\documentclass[aps,prl,superscriptaddress,amsfonts,amsmath,amssymb,floatfix,reprint,longbibliography]{revtex4-2}

\usepackage{url}
\usepackage{bm}
\usepackage{graphicx}
\usepackage{amsmath}
\usepackage{amstext}
\usepackage{amssymb}
\usepackage{amsfonts}
\usepackage{amsbsy}
\usepackage{verbatim}
\usepackage{color}
\usepackage[colorlinks=true, urlcolor=blue, linkcolor=blue, citecolor=blue, pdftex]{hyperref}
\usepackage{multirow}
\usepackage{floatrow}
\usepackage{float}
\usepackage{tikz}
\usepackage{gensymb}
\usepackage{textcomp}
\usepackage{enumitem}
\usepackage[version=3]{mhchem}
\usepackage{afterpage}
\usepackage{pbox}
\usepackage{makecell}
\usepackage{dsfont}
\usepackage{siunitx}
\usepackage{fancyhdr}
\usepackage{stackengine,wasysym,scalerel}
\usepackage{latexsym}
\usepackage{cancel}
\usepackage{array, multirow, bigdelim, makecell, booktabs}
\usepackage[misc]{ifsym}
\usepackage[normalem]{ulem}
\usepackage{times}

\definecolor{darkblue}{HTML}{004D6B}
\definecolor{darkred}{HTML}{8c1515}
\definecolor{darkgreen}{HTML}{006400}
\usepackage{hyperref}
\hypersetup{
	colorlinks=true,
	urlcolor=darkred,
	citecolor=darkblue,
	linkcolor=darkred,
	breaklinks
}

\begin{document}

\title{Pinch-points to half-moons and up in the stars: the kagome skymap}

\author{Dominik Kiese}
\thanks{These authors contributed equally to this work.}
\affiliation{Institute for Theoretical Physics, University of Cologne, 50937 Cologne, Germany}
\author{Francesco Ferrari}
\thanks{These authors contributed equally to this work.}
\affiliation{Institut f\"{u}r Theoretische Physik, Goethe-Universit\"{a}t Frankfurt, Max-von-Laue-Stra{\ss}e 1, D-60438 Frankfurt am Main, Germany}
\author{Nikita Astrakhantsev}
\thanks{These authors contributed equally to this work.}
\affiliation{Department of Physics, University of Z\"urich, Winterthurerstrasse 190, 8057 Z\"urich, Switzerland}
\author{Nils Niggemann}
\thanks{These authors contributed equally to this work.}
\affiliation{Dahlem Center for Complex Quantum Systems and Fachbereich Physik, Freie Universit{\"a}t Berlin, D-14195 Berlin, Germany}
\affiliation{Helmholtz-Zentrum Berlin f{\"u}r Materialien und Energie, D-14109 Berlin, Germany}
\affiliation{Department of Physics and Quantum Centers in Diamond and Emerging Materials (QuCenDiEM) group, Indian Institute of Technology Madras, Chennai 600036, India}
\author{Pratyay Ghosh}
\thanks{These authors contributed equally to this work.}
\affiliation{Institute for Theoretical Physics and Astrophysics, Julius-Maximilian's University of W\"urzburg, Am Hubland, D-97074 W\"urzburg, Germany}
\affiliation{Department of Physics and Quantum Centers in Diamond and Emerging Materials (QuCenDiEM) group, Indian Institute of Technology Madras, Chennai 600036, India}
\author{Tobias M\"uller}
\affiliation{Institute for Theoretical Physics and Astrophysics, Julius-Maximilian's
University of W\"urzburg, Am Hubland, D-97074 W\"urzburg, Germany}
\author{Ronny Thomale}
\affiliation{Department of Physics and Quantum Centers in Diamond and Emerging Materials (QuCenDiEM) group, Indian Institute of Technology Madras, Chennai 600036, India}
\affiliation{Institute for Theoretical Physics and Astrophysics, Julius-Maximilian's
University of W\"urzburg, Am Hubland, D-97074 W\"urzburg, Germany}
\author{Titus Neupert} 
\affiliation{Department of Physics, University of Z\"urich, Winterthurerstrasse 190, 8057 Z\"urich, Switzerland}
\author{Johannes Reuther}
\affiliation{Dahlem Center for Complex Quantum Systems and Fachbereich Physik, Freie Universit{\"a}t Berlin, D-14195 Berlin, Germany}
\affiliation{Helmholtz-Zentrum Berlin f{\"u}r Materialien und Energie, D-14109 Berlin, Germany}
\affiliation{Department of Physics and Quantum Centers in Diamond and Emerging Materials (QuCenDiEM) group, Indian Institute of Technology Madras, Chennai 600036, India}
\author{Michel J. P. Gingras}
\affiliation{Department of Physics and Astronomy, University of Waterloo, Waterloo, Ontario, Canada N2L 3G1}
\author{Simon Trebst}
\affiliation{Institute for Theoretical Physics, University of Cologne, 50937 Cologne, Germany}
\author{Yasir Iqbal}
\email{yiqbal@physics.iitm.ac.in}
\affiliation{Department of Physics and Quantum Centers in Diamond and Emerging Materials (QuCenDiEM) group, Indian Institute of Technology Madras, Chennai 600036, India}

\date{\today}


\begin{abstract}
    Pinch point singularities, associated with flat band magnetic excitations, are tell-tale signatures of Coulomb spin liquids. 
    While their properties in the presence of quantum fluctuations have been widely studied, the fate of the complementary non-analytic features -- shaped as half-moons and stars -- arising from adjacent shallow dispersive bands has remained unexplored. Here, we address this question for the spin $S=1/2$ Heisenberg antiferromagnet on the kagome lattice with second and third neighbor couplings, which allows one to tune the classical ground state from flat bands to being governed by shallow dispersive bands for intermediate coupling strengths. Employing the complementary strengths of variational Monte Carlo, pseudo-fermion functional renormalization group, and density-matrix renormalization group, we establish the quantum phase diagram. The U(1) Dirac spin liquid ground state of the nearest-neighbor antiferromagnet remains remarkably robust till intermediate coupling strengths when it transitions into a pinwheel valence bond crystal displaying signatures of half-moons in its structure factor. Our work thus identifies a microscopic setting that realizes one of the proximate orders of the Dirac spin liquid identified in a recent work [Song, Wang, Vishwanath, He, \href{https://doi.org/10.1038/s41467-019-11727-3}{Nat.\ Commun.\ {\bf 10}, 4254 (2019)}]. For larger couplings, we obtain a collinear magnetically ordered ground state characterized by star-like patterns.
\end{abstract}

\maketitle


\begin{figure*}[t]
    \centering
    \includegraphics[width=.95\columnwidth]{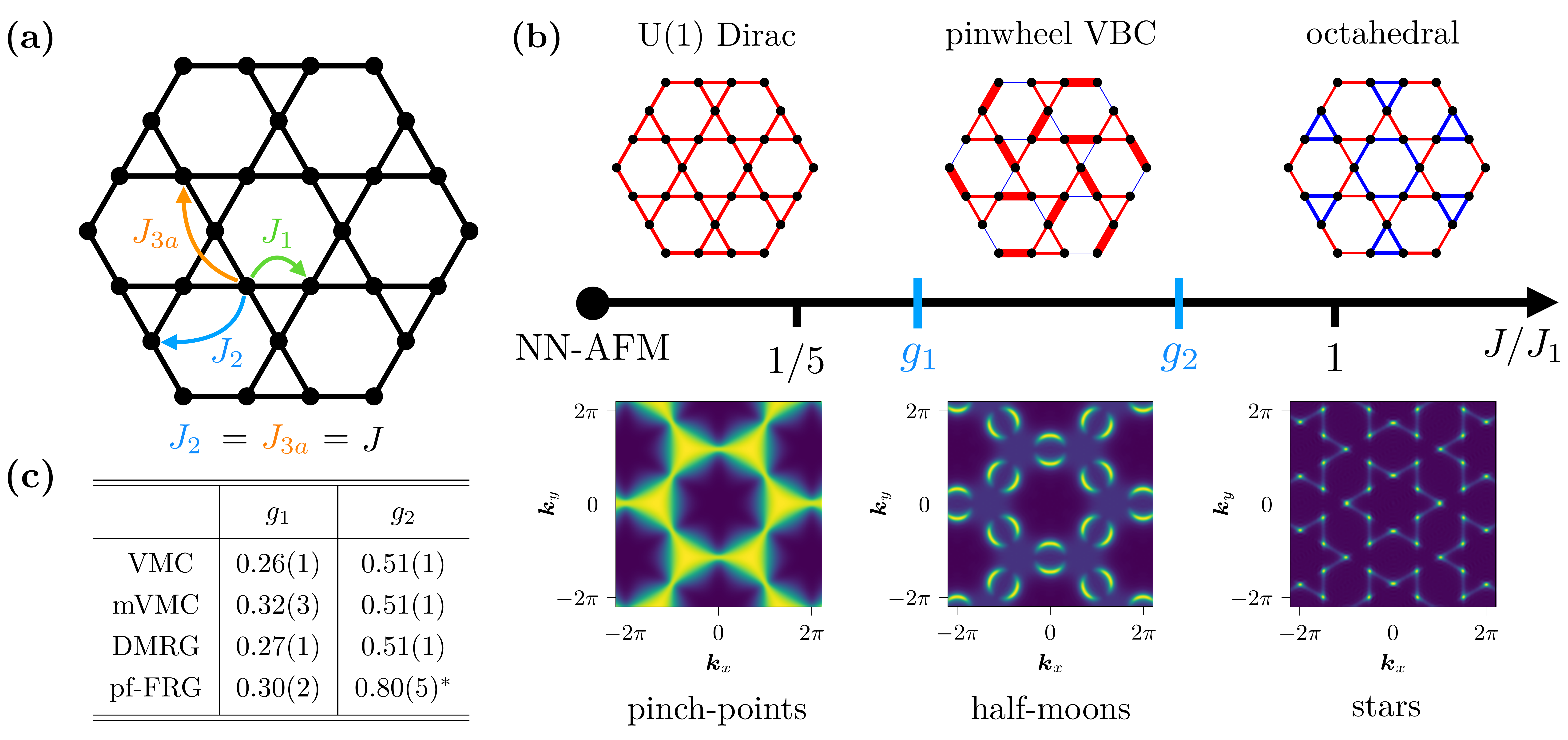}
    \caption{{\bf The kagome skymap.} (a) Illustration of first- ($J_1$), second- ($J_2$), and third-neighbor interactions along edges ($J_{3a}$) of the kagome lattice for the considered model. (b) The $S=1/2$ quantum phase diagram with (top panel) representative real-space spin-spin correlation profiles, with red (blue) bonds denoting antiferromagnetic (ferromagnetic) correlations, and (lower panel) spin structure factors of the different phases evaluated at $J/J_{1}=0.1$ (DSL), $J/J_{1}=0.4$ (pinwheel VBC), and $J/J_{1}=0.9$ (collinear order) from pf-FRG. (c) estimates of the phase boundaries ($g_{1}$ and $g_{2}$) obtained from the various approaches  employed in this work. While we see  agreement, within error bars, for $g_{1}$ for all approaches, the pf-FRG result for $g_2$ (marked by an asterisk), shows a notable deviation whose origin we discuss in Appendix~\ref{appendix:stars_pffrg}.}
    \label{fig:fig_1}
\end{figure*}


Classical spin models which admit a completion of squares belong to the distinct {\it genre} of ``maximally frustrated'' Hamiltonians which feature an exponentially large degenerate ground-state manifold~\cite{Chalker-1992,Ritchey-1993}. In two spatial dimensions, a celebrated example is the classical nearest-neighbor Heisenberg antiferromagnet (NNHAF) on the kagome lattice
\begin{equation}
    \mathcal{H}=J_{1}\sum_{\langle ij\rangle}\mathbf{S}_{i}\cdot \mathbf{S}_{j}=\frac{J_{1}}{2}\sum_{\bigtriangleup,\bigtriangledown}(\mathbf{S}_{1}+\mathbf{S}_{2}+\mathbf{S}_{3})^2 - J_{1}N \label{eqn:heisenberg}
\end{equation}
with $|\mathbf{S}_{i}|=1$ and $N$ the total number of spins. By virtue of the right-hand-side of Eq.~\eqref{eqn:heisenberg}, any spin configuration which satisfies $({\mathbf S}_{1}+{\mathbf S}_{2}+{\mathbf S}_{3})=0$ on each triangle qualifies as a classical ground state.
The emergence of such a local constraint leads to the formation of a Coulomb spin liquid~\cite{Henley-2010},
with algebraically decaying spin-spin correlations in real space, which gives structure to the exponentially large manifold of degenerate ground states. In Fourier space, these correlations most strikingly manifest themselves in the presence of non-analytic features in the structure factor called {\em pinch points}~\cite{Garanin-1999,Zhitomirsky-2008}. Remarkably, this classical Coulomb phase remains stable~\cite{Mizoguchi-2018,Li-2021} even in the presence of additional couplings along a fine-tuned line when second neighbor ($J_{2}$) and third neighbor along edges ($J_{3a}$) [see Fig.~\ref{fig:fig_1}(a)] 
are concurrently introduced and of equal strength, i.e. $J_{2}=J_{3a}$ ($\equiv J$ henceforth). This can be readily understood when diagonalizing the spin exchange Hamiltonian $\mathcal{H}(\mathbf{k})$ in momentum space~\cite{Luttinger1951,Luttinger1946,Lyons1960}, which reveals that the characteristic flat band of the NNHAF persists~\cite{Mizoguchi-2018} up to $J/J_{1}=1/5$. For $J/J_{1}>1/5$, a shallow dispersive band starts to cut below the flat band in parts of the Brillouin zone~\cite{Mizoguchi-2018}, which in turn gives rise to pairs of {\em half-moons}, i.e., crescent shaped arcs in the static structure factor~\cite{Robert-2008}, with the flat band remaining close-by with a multitude of low-energy excitations \footnote{Half-moons also appear for Ising models on the kagome and pyrochlore lattices for $0<J/J_{1}<1/3$~\cite{Mizoguchi-2017} and $0<J/J_{1}<1/4$~\cite{Rau-2016,Udagawa-2016}, respectively.}. On a deeper level, the formation of half-moons in the static structure factor results from a non-analyticity in the dispersive-band eigenvectors as a function of momentum and, given the completeness of the eigenvector basis, can be viewed as necessarily arising in order to complement the singularity in the momentum dependence of the flat-band eigenvectors ~\cite{Mizoguchi-2018,Yan-2018}. With increasing $J/J_{1}$, the radius of the half-moon continuously grows and at $J/J_{1}=1$, the half-moons from different Brillouin zones coalesce, giving rise to a {\em star} pattern in the static structure factor. While in the case of Ising spins, which show a similar sequence of momentum space signatures as a function of $J/J_{1}$, the nature of the half-moons and star phases has a well-understood real-space picture in terms of magnetic clustering of topological charges~\cite{Mizoguchi-2017,Rau-2016,Udagawa-2016}, for continuous (Heisenberg) spins, the nature of the real-space clustering and its freedom to continuously evolve with $J$ is far more involved and not yet completely understood~\cite{Mizoguchi-2018}. 

Much of the interest in the kagome {\sl quantum} antiferromagnet as a potential host to highly entangled quantum states owes its origin to the realization that its classical ground state is governed by flat bands -- an opportunity for otherwise residual quantum effects to dictate the macroscopic ground state. Thence, tuning the pairwise exchange along the maximally frustrated axis $(J_{2}=J_{3a}\equiv J)$ which, classically, is tuned to have a flat band over an extended region in parameter space, should provide a fertile playground to potentially realize novel states of matter also in the quantum model. For one, the U(1) Dirac spin liquid (DSL)~\cite{Hastings-2000,Hermele-2008,Song-2019} ground state of the NNHAF~\cite{Iqbal-2013,Iqbal-2014_gap,Iqbal-2011Z2,He-2017} is indeed known to be fragile to magnetic order when perturbed by longer-range Heisenberg couplings~\cite{Iqbal-2021j1j2,Iqbal-2015j1j2} or Dzyaloshinskii-Moriya interactions~\cite{Lee-2018}, as expected for algebraic spin liquids, but its fate along the maximally frustrated direction of interest here is unknown. In particular, this parameter axis may afford a higher degree of stability to the U(1) DSL against long-range order, and one may wonder whether the DSL naturally gives way to other exotic quantum phases as one marches along this direction. On a conceptual level, instabilities of the DSL have recently been rigorously classified~\cite{Song-2019} in field theoretical work. But it remains an open challenge to identify microscopic settings in which these instabilities manifest themselves and what tell-tale signatures they come along with that might be accessible in experimental studies.

In this manuscript, we take an important step in this direction by establishing the quantum counterpart to the classical half-moon phase as a pinwheel valence bond crystal state which the DSL transitions into only for finite coupling strength. We do so by employing complementary numerical quantum many-body approaches to build a detailed picture of the $S=1/2$ quantum phase diagram along the maximally frustrated axis for $J/J_{1}>0$, resolving the characteristic real-space and Fourier-space signatures of all quantum phases. The numerical approaches include fermionic variational Monte Carlo (VMC) with versatile Gutzwiller projected Jastrow wave functions~\cite{becca_quantum_2017}, many-variable variational Monte Carlo (mVMC) with unconstrained optimization of the Bardeen-Cooper-Schrieffer (BCS) pairing function (supplemented with symmetry projectors)~\cite{misawa2019mvmc,doi:10.1143/JPSJ.77.114701}, one-loop pseudo-fermion functional renormalization group (pf-FRG)~\cite{ReutherOrig}, and density-matrix renormalization group (DMRG)~\cite{White-1992}. The resulting quantum phase diagram is shown in Fig.~\ref{fig:fig_1}, where cumulative and complementary evidence from all employed approaches shows that the ground state remains nonmagnetic over an appreciably wide span of parameter space [see Fig.~\ref{fig:fig_1}(b)], notably extending far beyond the classical domain ($0\leqslant J/J_{1}\leqslant0.2$) where flat bands are lowest in energy. This nonmagnetic region is composed of two phases: (i) the U(1) Dirac spin liquid (DSL) for $0\leqslant J/J_{1}\lesssim 0.26$ characterized by soft maxima at the pinch points in its spin structure factor $\chi(\mathbf{k})$, 
and (ii) a 12-site unit cell, $C_{6}$ symmetric pinwheel valence bond crystal (VBC) for $0.26 \lesssim J/J_{1}\lesssim 0.51$, displaying signatures of half-moons in $\chi(\mathbf{k})$, see Fig.~\ref{fig:fig_1}(b). Our analysis indicates the DSL-VBC transition to be first-order as ascertained on finite systems from a sudden change in the spin-spin correlation profile and a crossing of the energies. For $J/J_{1}\gtrsim 0.51$, the VBC gives way, via a first-order transition, to collinear long-range magnetic order~\cite{Grison-2020,Messio-2011} with signatures of a star-like pattern in  $\chi(\mathbf{k})$. 


\begin{figure}
    \centering
    \includegraphics[width=\columnwidth]{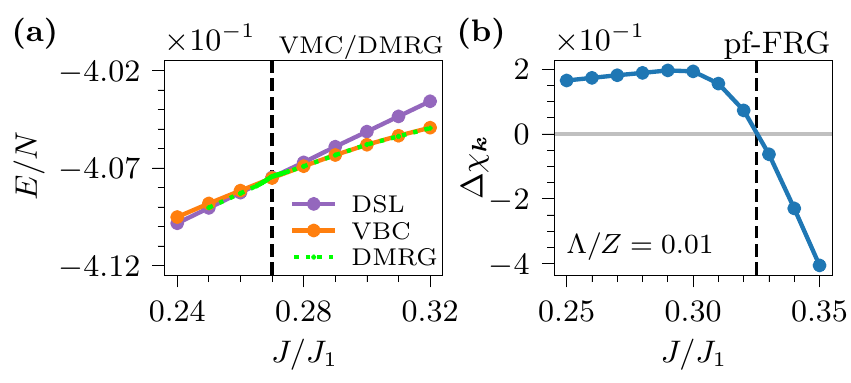}
    \caption{{\bf Transition into half-moon phase.} (a) From VMC, the evolution with $J/J_{1}$ of the energy per site of the DSL and VBC states ($3 \times 12 \times 12$ lattice). DMRG energies are also shown for comparison. (b) From pf-FRG, the variation of the spectral measure $\Delta\chi_{\boldsymbol{k}}$ (see text below) with $J/J_{1}$ evaluated at the lowest simulated RG cutoff $\Lambda/Z=0.01$ where $Z=\sqrt{J_{1}^{2}+2J^{2}}$.}
    \label{fig:first_transition}
\end{figure}

{\it Results}.
We set the stage, by observing that across our numerical approaches we find that the ground state energy is seen to {\em increase} with $J/J_{1}$, reflecting an enhanced degree of frustration at variance with conventional expectation that the NNHAF represents the point of maximal frustration, which is relieved upon inclusion of long-range couplings. The presence of a pronounced kink in the evolution of the ground state energy is indicative of a phase transition [see Fig.~\ref{fig:first_transition}(a)] which we estimate to be at $g_{1}=0.27(1)$ via an analysis of its derivative (from our DMRG calculations). This value is also corroborated by the behavior of the von Neumann entanglement entropy which starts decreasing sharply at $g_{1}$ [see Fig.~\ref{fig:entropy}] indicating the formation of a less entangled state.

To probe the nature of the ensuing states, let us start by discussing results from our fermionic VMC approach with versatile Gutzwiller-projected wave functions constructed in a manner enabling us to accurately study the competition between nonmagnetic quantum spin liquid (QSL) and VBC phases, together with magnetically ordered states. Such a unified framework has been met with success in its application to a wide range of frustrated spin models~\cite{Iqbal-2016tri,Iqbal-2021j1j2,Iqbal-2021shuriken,Iqbal-2018breathing,Iqbal-2015j1j2}. Our calculations are performed on $3\times L\times L$ clusters respecting the full symmetry of the kagome lattice. For the $S=1/2$ NNHAF, there is emerging consensus towards a U(1) DSL ground state~\cite{Ran-2007,Iqbal-2013,He-2017,Zhu-2018}, which is known to yield the lowest variational energy~\cite{Ran-2007,Iqbal-2013}. Upon including a $J$ coupling, we investigate for the potential instability of the U(1) DSL to symmetric $\mathds{Z}_{2}$~\cite{Lu-2011}, chiral U(1)~\cite{Bieri-2015}, chiral $\mathds{Z}_{2}$~\cite{Bieri-2016}, and lattice nematic $\mathds{Z}_{2}$~\cite{Schaffer-2017} QSLs. We also probe for possible dimerization tendencies into VBCs with various unit cell sizes up to 36 sites and different symmetries~\cite{Iqbal_2012,Iqbal-2011a,Hermele-2008,Singh-2007,Hastings-2000}. Our analysis finds a remarkable robustness of the U(1) DSL to the above-mentioned potential instabilities over a wide range along the maximally frustrated axis extending up till $J/J_{1}=0.26(1)$, which we note is beyond the range of $J/J_{1}$ for the classical model where the flat band is the lowest in energy~\cite{Mizoguchi-2018}. 

At $J/J_{1}=0.26(1)$, we detect a {\em dimer instability} of the DSL towards a VBC ground state in our VMC calculations. This VBC state is found to be characterized by a $2\times2$ enlarged unit cell with a $C_{6}$ invariant {\it pinwheel} structure of spin-spin correlations in real space which breaks reflection symmetries [see Fig.~\ref{fig:fig_1}(b)]. The formation of such a VBC state is further corroborated by an enhanced dimer response (see Fig.~\ref{eq:DimerResponse}). Interestingly, such a pattern of strong/weak bonds has previously been identified as descending from confinement transitions of $\mathds{Z}_{2}$ spin liquids~\cite{Huh-2011} (left panel of Fig.~1 therein), and recently proposed in Ref.~\cite{Song-2019} [Fig.~3(c) therein] as a potential instability of the U(1) DSL resulting from a condensation of a $C_{6}$ invariant mass and the associated monopole terms. 
Our finding of a $C_{6}$ symmetric VBC, as opposed to other less symmetric patterns [Fig.~2(c) in Ref.~\cite{Yan-2011}], is likely connected to the fact that the imaginary expectation value of the monopole condensation responsible for this reflection symmetry breaking pattern also optimizes the Landau potential~\cite{Song-2019}. It is worth pointing out that our VBC pattern is distinct from the $2\times2$ enlarged VBC patterns previously proposed in Fig.~4 of Ref.~\cite{Hastings-2000} and Fig.~5 of Ref.~\cite{Hermele-2008} which do not break reflections (though these pattern also minimize the Landau potential as noted in Ref.~\cite{Song-2019}). While, the DSL to VBC transition is allowed to be continuous, our microscopic calculations find it to be first-order as inferred from a level-crossing of the energies of the two states [see Fig.~\ref{fig:first_transition}(a)] together with the observation of an abrupt change in the nearest-neighbor spin-spin correlation profile. We show that the energy gain of the VBC w.\,r.\,t.\ the U(1) DSL is non-zero for $J/J_{1}>0.26(1)$ and remains so on all finite size systems we simulated, indicating size-consistency of the VBC state and its stability in the thermodynamic limit. 

Further support for the pinwheel VBC state comes from mVMC calculations at $J/J_{1} = 0.4$, for which we measure the real-space dimer-dimer correlation pattern (see Fig.\,\ref{fig:dimer_real_space_pattern}) where the emergence of the $C_{6}$ symmetric pinwheel VBC is also manifest. We also construct a symmetry-breaking dimer operator with non-vanishing susceptibility extrapolated to the thermodynamic limit (see Fig.\,\ref{fig:dimer_correlations_pattern}). An analysis of the latter suggests a triply-degenerate $C_3$-related order parameter, with the three $M$-points momenta setting the spatial dependence, which signals a VBC behavior with the spontaneous $C_3$-symmetry breaking. However, the equal-weight sum of these three basis functions of the dominant irreducible representations results into an effective $C_6$ symmetric pinwheel pattern as obtained within VMC [see Fig.\,\ref{fig:fig_1}\,(b)], which we illustrate in the inset of Fig.\,\ref{fig:dimer_correlations_pattern}. The corresponding susceptibility decreases rapidly as $J/J_{1} \to 0$, substantiating a transition to a quantum spin liquid phase from the VBC.

\begin{figure}[t]
    \centering
    \includegraphics[width=.85\columnwidth]{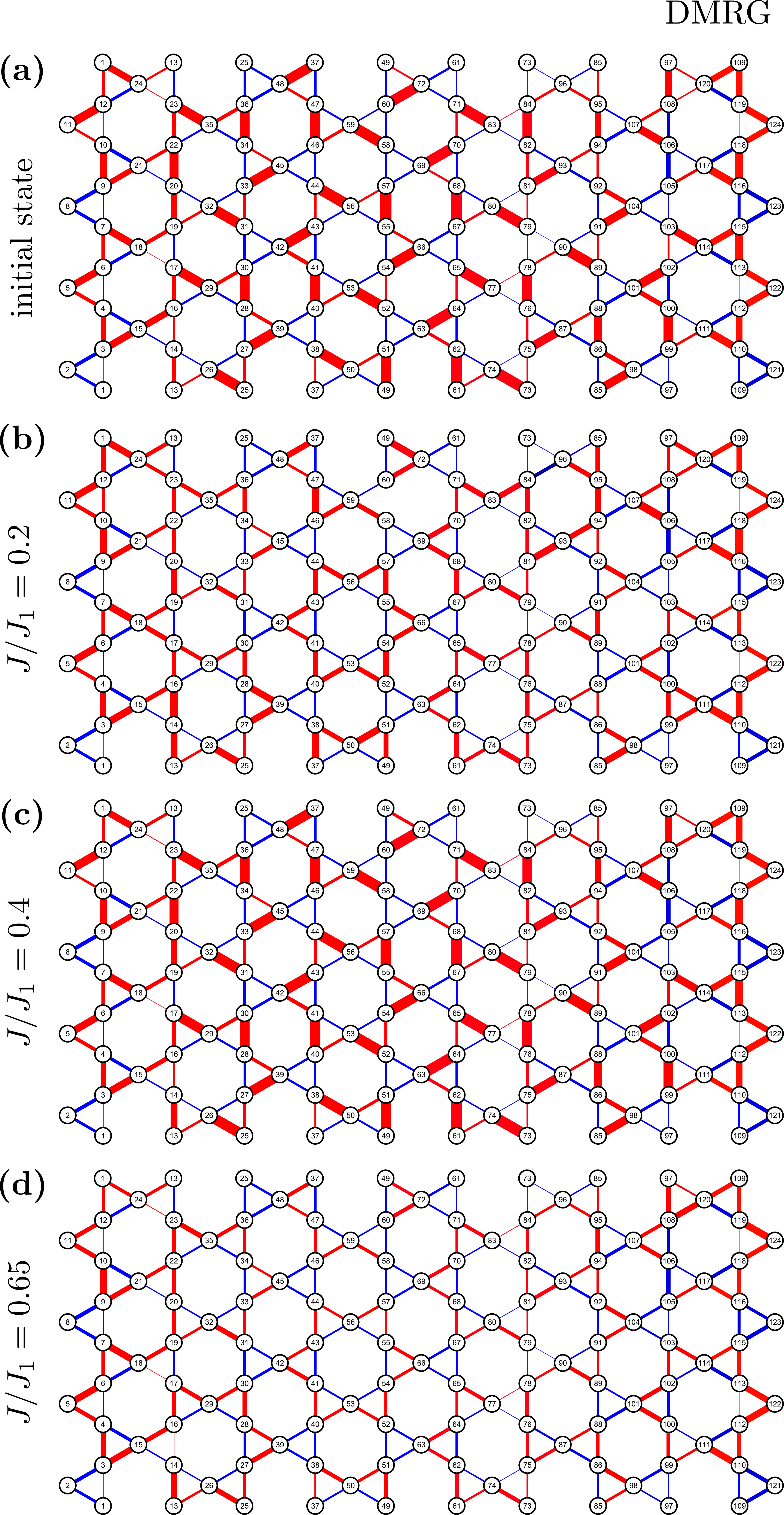}
    \caption{{\bf Nearest-neighbor spin-spin correlations} obtained from DMRG.
    		(a) the initial state obtained at $J/J_1=0.4$ with a bias in the Hamiltonian ($5\%$ of $J_{1}$) that favours the onset of the pinwheel VBC. The final converged states obtained after 24 sweeps at (b) $J/J_1=0.2$, (c) $J/J_1=0.4$ and (d) $J/J_1=0.65$.}
    \label{fig:DMRG_pattern_2}
\end{figure}

To probe the aforementioned VBC order within DMRG, we start by imposing the pinwheel VBC pattern  (via small pinning fields) in a trial wavefunction that is then used as initial state for subsequent DMRG calculations performed with the original unperturbed Hamiltonian deep within the three phases of interest, namely, at $J/J_1=0.2$, $J/J_1 = 0.4$, and $J/J_1=0.65$. This procedure allows us to probe the stability of the initial pinwheel VBC state for these three phases or, alternatively, see its melting into different quantum states. We see that for $J/J_{1}=0.4$ [see Fig.~\ref{fig:DMRG_pattern_2}(c)], the removal of the bias hardly affects the initial state thus providing strong support for the pinwheel VBC as true ground state in this regime. This is further corroborated by the fact that at $J/J_{1}=0.2$ and $0.65$, the VBC pattern is progressively washed out [see Fig.~\ref{fig:DMRG_pattern_2}(b) and see Fig.~\ref{fig:DMRG_pattern_2}(d)]. Together, these results provide a smoking gun signature for the formation of the pinwheel VBC state in the range $J\in(g_1,g_2)$ [see Fig.~\ref{fig:fig_1}(c)]. 

\begin{figure}[t]
    \centering
    \includegraphics[width=\columnwidth]{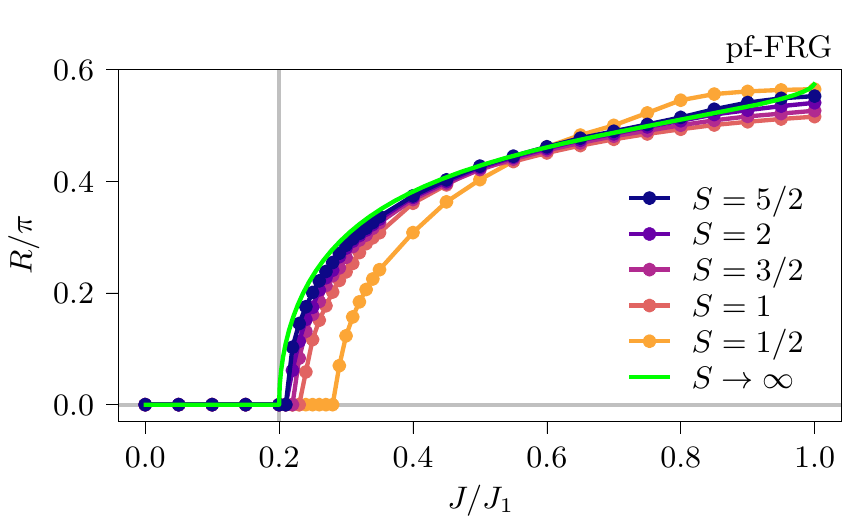}
    \caption{{\bf Half-moon radii.} From pf-FRG, we show for different values of spin-$S$~\cite{LargeS}, the evolution with $J/J_{1}$ of the radius of the half-moons characterizing the pinwheel VBC. The large-$S$ (classical) result is from Ref.~\cite{Mizoguchi-2018}.}
    \label{fig:radii}
\end{figure}

In Fourier space, the hallmark of the onset of the VBC order, as obtained within pf-FRG, is the splitting of the pinch points (M-points of the extended Brillouin zone), where the maxima of $\chi(\mathbf{k})$ are located for the DSL, into two symmetric half-moons resulting in the maxima of the intensity now being located at generic $(0,q_{y})$ (and symmetry related) incommensurate points, as captured in an earlier pf-FRG study of the same model~\cite{Buessen2016}. Given that the DSL and VBC phases can also be distinguished by comparing $\chi(\mathbf{k})$ along two cuts in momentum space, i.e., $\Gamma -K$ and $\Gamma -M$ segments, more precisely, we define a ``spectral measure'' $\Delta\chi_{\boldsymbol{k}}$ as the difference between the maxima along these two cuts, i.e., $\Delta\chi_{\boldsymbol{k}}=\chi^{\rm max}(\mathbf{k}\in\Gamma -K)-\chi^{\rm max}(\mathbf{k}\in\Gamma -M)$. The splitting of the pinch point into half-moons correspond to a downturn in the value of $\Delta\chi$ while the zero crossing of $\Delta\chi$ indicates that the half-moons become the dominant feature in $\chi(\mathbf{k})$. Based on these two signatures,  we estimate the onset of VBC from pf-FRG at $J/J_{1}=0.30(2)$ [see Fig.~\ref{fig:first_transition}(b) and Fig.~\ref{fig:FRG_spectral}], in good agreement with the other employed approaches. The evolution of the radius of the half-moon as a function of $J/J_{1}$ obtained from pf-FRG is shown in Fig.~\ref{fig:radii}, where for $S=1/2$ one observes an appreciable deviation from the reported large-$S$ result~\cite{Mizoguchi-2018}. For progressively increasing values of $S$, the known large-$S$ behavior~\cite{Mizoguchi-2018} is approached. Within the VMC calculation, the splitting of the pinch point maxima into half-moons is observed deep inside the VBC phase as shown in Fig.~\ref{fig:halfmoon_vmc}. Similarly, deep inside the VBC phase, the $\chi(\mathbf{k})$  obtained from mVMC shows maxima at incommensurate $(0,k_{y})$ points as shown in Fig.~\ref{fig:magnetic_2d}.

\begin{figure}
    \centering
    \includegraphics[width=\columnwidth]{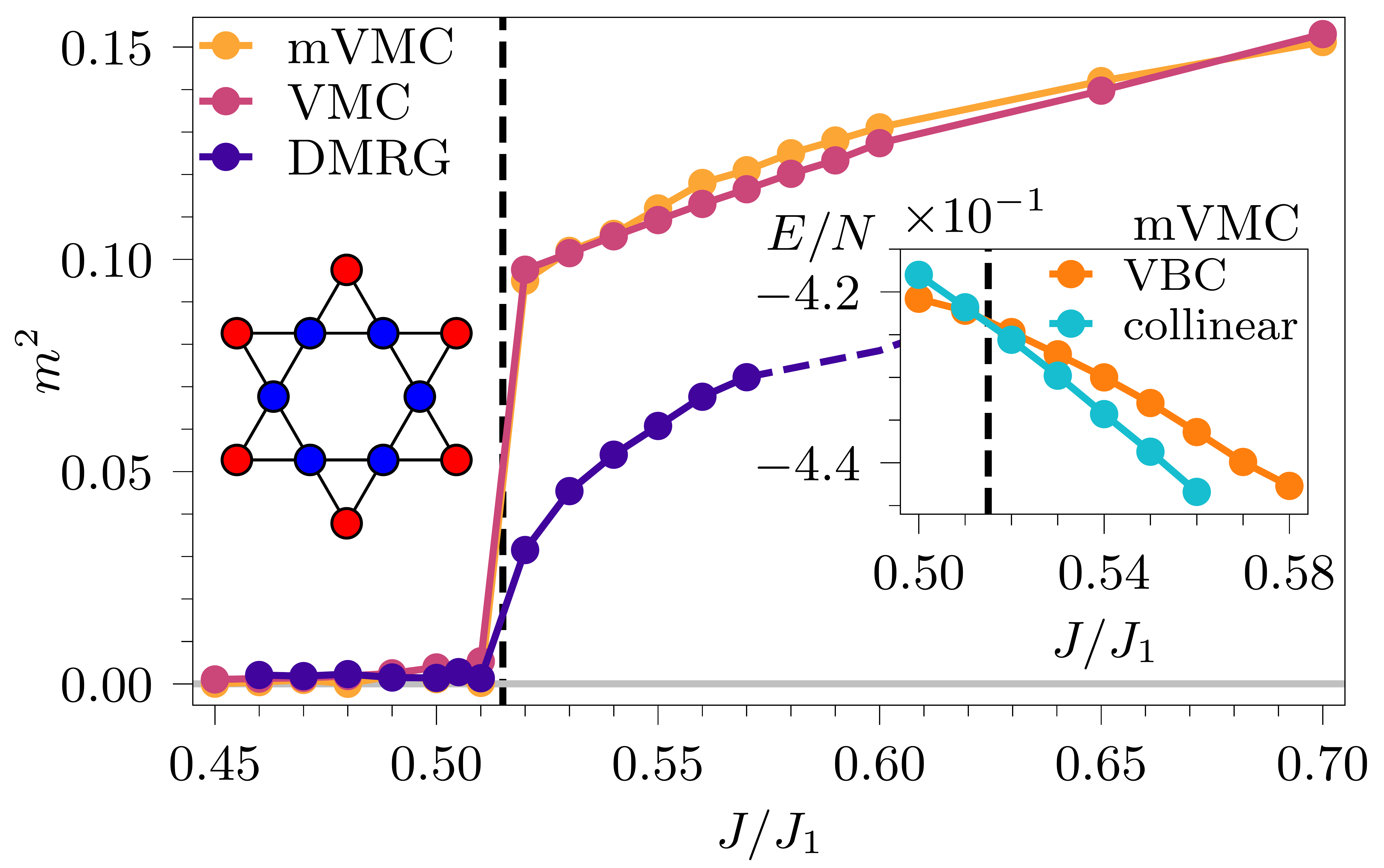}
    \caption{{\bf Transition into the star phase.} The behavior of the square of the sublattice magnetization $m^{2}$ with $J/J_{1}$ near the transition from the pinwheel VBC into collinear magnetic order [illustrated in the inset, with blue and red spins pointing in opposite directions]. The results from VMC and mVMC are for a $3\times8\times8$ site cluster [see Fig.~\ref{fig:m2_vmc_scaling} for finite-size scaling results of $m^{2}$ from VMC to the thermodynamic limit], while those from DMRG are obtained on a YC8-8 cylinder.}
    \label{fig:mvmc_m2}
\end{figure}

Finally, let us turn to the transition into the star phase. To this end,  we show, in Fig.~\ref{fig:mvmc_m2}, the evolution of the square of the sublattice magnetization $m^{2}$ with $J/J_{1}$, as obtained from mVMC, VMC, and DMRG. One observes a sudden change to a finite value of $m^{2}$ for $J/J_{1}>0.51(1)$, indicating the onset of long-range collinear spin order with a 12-site magnetic unit cell (see inset of Fig.~\ref{fig:mvmc_m2})~\cite{Grison-2020}. While the estimate of the phase boundary from these three approaches shows  remarkable agreement, the comparatively smaller values of $m^2$ inside the ordered phase obtained in DMRG can be ascribed to the quasi one-dimensional character of the cylindrical geometries. The abrupt nature of the jump in the value of $m^{2}$ observed in mVMC and VMC, together with the crossing of the energies of the disordered VBC and magnetically ordered states across the transition point (see inset of Fig.~\ref{fig:mvmc_m2}), lends evidence in favor of a first-order character of the transition. Similar conclusions are drawn from VMC via  finite-size scaling of $m^{2}$ for different values of $J/J_{1}$ (see Fig.~\ref{fig:m2_vmc_scaling}), wherein one observes a jump in the value of $m^{2}$ in the thermodynamic limit. The collinear magnetically ordered state displays a star-like pattern of intensity distribution in $\chi(\mathbf{k})$  [see Fig.~\ref{fig:fig_1}(b)] with maxima at the location expected for the octahedral regular magnetic order~\cite{Messio-2011}. It is worth noting that for $S=1/2$ the phase boundary between the half-moon and star phases considerably shifts to a smaller value of $J/J_{1}=0.51(1)$, compared to the classical boundary at $J/J_{1}=1$, highlighting significant effects of quantum fluctuations.


{\it Discussion}.
Moving the ground state of the kagome antiferromagnet along the maximally frustrated line is a complicated endeavor -- as such it
is quite fulfilling to see the remarkable agreement between our complementary numerical approaches yielding a consistent understanding of momentum and 
real space signatures of the ground state phases and their respective boundaries; a feat that would not have been imaginable only a few years ago. One might hope that the U(1) DSL, half-moon, and star phases will have a window of stability away from the maximally frustrated axis. It would thus be of interest to search and identify materials promising to realize the Dirac spin liquid phase and which lie within this region of stability. The recently studied material \ce{YCu3(OH)6Br2 [Br_$x$(OH)_{$1-x$}]}~\cite{Zeng-2022} wherein signatures of DSL behavior has been presented, could serve as a potential material candidate warranting further investigation. Another interesting candidate material might be the distorted kagome compound \ce{Rb2Cu3SnF12}~\cite{Matan-2010} where indications for a pinwheel VBC have been reported. One may be able to approach the maximally frustrated line by effectively varying the super-exchange couplings by application of hydrostatic or uniaxial pressure to vary the super-exchange bond angles~\cite{Iqbal-2015_para}. On the theoretical front, given the persistent and enhanced frustration upon inclusion of $J$, it would be interesting to ascertain the extent of the nonmagnetic phase of the spin $S=1$ NNHAF, and decipher the corresponding real-space nature of the half-moon phase. Finally, it would be worth exploring the corresponding quantum phase diagram on the pyrochlore lattice, which similarly at the classical level is host to persistent flat bands, as well as half-moon and star phases~\cite{Mizoguchi-2018,Rau-2016}.

During completion of this manuscript, we were made aware of a paper by Lugan {\it et al}. studying the same model with a complementary bosonic method.


{\it Acknowledgments}. 
We thank Federico Becca, Subhro Bhattacharjee, Ludovic Jaubert, Harald Jeschke, Arnaud Ralko, and Arnab Sen for insightful discussions. 
D.\,K., N.\,N., J.\,R., and S.\,T acknowledge support from the Deutsche Forschungsgemeinschaft (DFG, German Research Foundation), within Project-ID 277101999 CRC 183 (Project A04).
F.\,F. acknowledges support from the Alexander von Humboldt Foundation through a postdoctoral Humboldt fellowship.
N.\,A. is funded by the Swiss National Science Foundation, grant number: PP00P2{\_}176877. The mVMC simulations were supported by the RSF grant (project No.\,21-12-00237). 
The work in Wurzburg was supported by the Deutsche Forschungsgemeinschaft (DFG, German Research Foundation) through Project-ID 258499086-SFB 1170 and the Wurzburg-Dresden Cluster of Excellence on Complexity and Topology in Quantum Matter – ct.qmat Project-ID 390858490-EXC 2147.
M.J.P.G is supported by the NSERC of Canada and the Canada Research Chair program (M. J. P. G., Tier 1).
Y.\,I. acknowledges financial support by the Science and Engineering Research Board (SERB), Department of Science and Technology (DST), India through the Startup Research Grant No.~SRG/2019/000056, MATRICS Grant No.~MTR/2019/001042, and the Indo-French Centre for the Promotion of Advanced Research (CEFIPRA) Project No. 64T3-1. This research was supported in part by the National Science Foundation under Grant No.~NSF~PHY-1748958, the Abdus Salam International Centre for Theoretical Physics (ICTP) through the Simons Associateship scheme funded by the Simons Foundation, IIT Madras through the Institute of Eminence (IoE) program for establishing the QuCenDiEM group (Project No. SB20210813PHMHRD002720) and FORG group (Project No. SB20210822PHMHRD008268), the International Centre for Theoretical Sciences (ICTS), Bengaluru, India during a visit for participating in the program “Novel phases of quantum matter” (Code: ICTS/topmatter2019/12). 
N.\,N. thanks IIT Madras for funding a three-month stay through an International Graduate Student Travel award which facilitated completion of this research work.
J.\,R. thanks IIT Madras for a Visiting Faculty Fellow position under the IoE program during which part of the research work and manuscript writing were carried out.
D.~K. and S.\,T. acknowledge usage of the JURECA Booster and JUWELS cluster at the Forschungszentrum Juelich and
the Noctua2 cluster at the Paderborn Center for Parallel Computing (PC$^2$).
N.\,N. and J.\,R. acknowledges the use of the CURTA cluster at FU Berlin~\cite{Bennett2020}.
N.\,A. acknowledges the usage of computing resources of the federal collective usage center ``Complex for simulation and data processing for mega-science facilities'' at NRC ``Kurchatov Institute''.
T.\,M., P.\,G., and R.\,T. gratefully acknowledge the Gauss Centre for Supercomputing e.\,V. 
for funding this project by providing computing time on the GCS Supercomputer SuperMUC at Leibniz Supercomputing Centre. 
Y.\,I. acknowledges the use of the computing resources at HPCE, IIT Madras. 



%


\newcommand{\beginsupplement}{%
        \setcounter{table}{0}
        \renewcommand{\thetable}{S\arabic{table}}
        \renewcommand{\theHtable}{S\arabic{table}}%
        \setcounter{figure}{0}
        \renewcommand{\thefigure}{S\arabic{figure}}%
        \renewcommand{\theHfigure}{S\arabic{figure}}
        \setcounter{equation}{0}
        \renewcommand{\theequation}{S\arabic{equation}}%
        \renewcommand{\theHequation}{S\arabic{equation}}%
        \setcounter{page}{1}
        \setcounter{secnumdepth}{3}
     }
 
\renewcommand*{\citenumfont}[1]{S#1}
\renewcommand*{\bibnumfmt}[1]{[S#1]}
 
\newcommand\blankpage{%
    \null
    \thispagestyle{empty}%
    \addtocounter{page}{-1}%
    \newpage
}
\newpage
\newpage
\chead{{\large \bf{--- Supplemental Material ---}}}
\thispagestyle{fancy}

\beginsupplement

\maketitle


\section{Pseudo-fermion functional renormalization group}

The pseudo-fermion functional renormalization group approach (pf-FRG)~\cite{ReutherOrig} approximates the original spin model by a fermionic Hamiltonian using an Abrikosov fermion representation 
\begin{align}
    S^{\mu}_i = \frac{1}{2} \sum_{\alpha, \beta} f^{\dagger}_{i, \alpha} \sigma^{\mu}_{\alpha, \beta} f^{\phantom{\dagger}}_{i, \beta} \,,
    \label{eq:partons}
\end{align}
of the spin operators together with the soft-constraint $\langle \sum_{\alpha} f^{\dagger}_{i,\alpha} f^{\phantom{\dagger}}_{i, \alpha} \rangle = 1$ on every lattice site. Fluctuations around this average decrease during the RG flow and can be further suppressed adding level repulsion terms to the Hamiltonian~\cite{thoenniss2020}, yet the qualitative results, especially with respect to the nature of the ground state, appear robust with respect to small variations of the number of particles per site~\cite{thoenniss2020, KieseSpinValley, BuessenDiamond, LargeS}. 

The flow equations are generated by implementing an infrared cutoff $\Lambda$ into the bare propagator $G_{0}(\omega) = (i\omega)^{-1}$ of the pseudo-fermion Hamiltonian and taking derivatives of one-particle irreducible vertices with respect to it. The resulting hierarchy of ordinary differential equations is not closed and thus needs to be truncated, usually by discarding all $n$-particle vertices with $n > 2$. In pf-FRG one needs to incorporate some contributions from the three-particle vertex by means of the so called Katanin truncation~\cite{ReutherOrig}, which feeds back the self-energy flow into the flow of the two-particle vertex. The corresponding flow equations for the self-energy $\Sigma$ and two-particle vertex $\Gamma$ then read
\begin{align}
	\frac{d}{d\Lambda} \Sigma^{\Lambda}(1) &= -\frac{1}{2\pi} \sum_2 \Gamma^{\Lambda}(1, 2| 1, 2) S^{\Lambda}(2) \notag \\
	\frac{d}{d\Lambda} \Gamma^{\Lambda}(1', 2'| 1, 2) &= -\frac{1}{2\pi} \sum_{3, 4} \big{[} \Gamma^{\Lambda}(3, 4| 1, 2) \Gamma^{\Lambda}(1', 2'| 3, 4) \notag  \\ 
	& - \Gamma^{\Lambda}(1', 4| 1, 3) \Gamma^{\Lambda}(3, 2'| 4, 2) - (3 \leftrightarrow 4)          \notag  \\ 
	& + \Gamma^{\Lambda}(2', 4| 1, 3) \Gamma^{\Lambda}(3, 1'| 4, 2) + (3 \leftrightarrow 4) \big{]}  \notag  \\
	& \times \partial_{\Lambda} (G^{\Lambda}(3) G^{\Lambda}(4)) \,,
	\label{eq:flow}
\end{align}
where $G^{\Lambda}$ denotes the full fermionic propagator and $S^{\Lambda} \equiv -\frac{d}{d\Lambda} G^{\Lambda}|_{\Sigma^{\Lambda} = \text{const.}}$ the single-scale propagator. Here, multi-indices $1 = (i_1, \alpha_1, \omega_1)$ comprise a lattice, spin and Matsubara frequency index.

To characterize the physical field theory that the pf-FRG is flowing towards, one usually computes spin-spin correlators 
\begin{align}
	\chi_{ij} = \chi^{zz}_{ij}(i \omega = 0) = \int_{0}^{\beta} d\tau \langle T_{\tau} S^{z}_{i}(\tau) S^{z}_{j}(0) \rangle \,,
	\label{eq:susceptibility}
\end{align}
from renormalized pseudo-fermion vertices (we suppress the $\Lambda$ dependence here for brevity) and checks whether long-range order manifests as an instability in their flow. The associated spin configuration can then be determined by Fourier transforming $\chi_{ij}$ to momentum space and locating the position of the incipient Bragg peaks. A paramagnetic phase, on the other hand, is signified by a smooth flow down to the infrared $\Lambda \to 0$ with broadened features in the structure factor $\chi(\boldsymbol{k})$.

We use the \texttt{PFFRGSolver.jl}~\cite{kiese2021,PFFRGSolver} software package to perform the integration of the flow equations in this manuscript. All calculations are performed on a $48 \times 36^2$ frequency grid with absolute error tolerances $a_{\mathrm{tol}} = 10^{-8}$ and a relative error tolerance $r_{\mathrm{tol}} = 10^{-2} \ (10^{-4})$ for the differential equation solver (Matsubara frequency integrals). The real-space truncation is set to $L = 24$ bonds away from the origin.


\begin{figure}[b]
    \centering
    \includegraphics[width=\columnwidth]{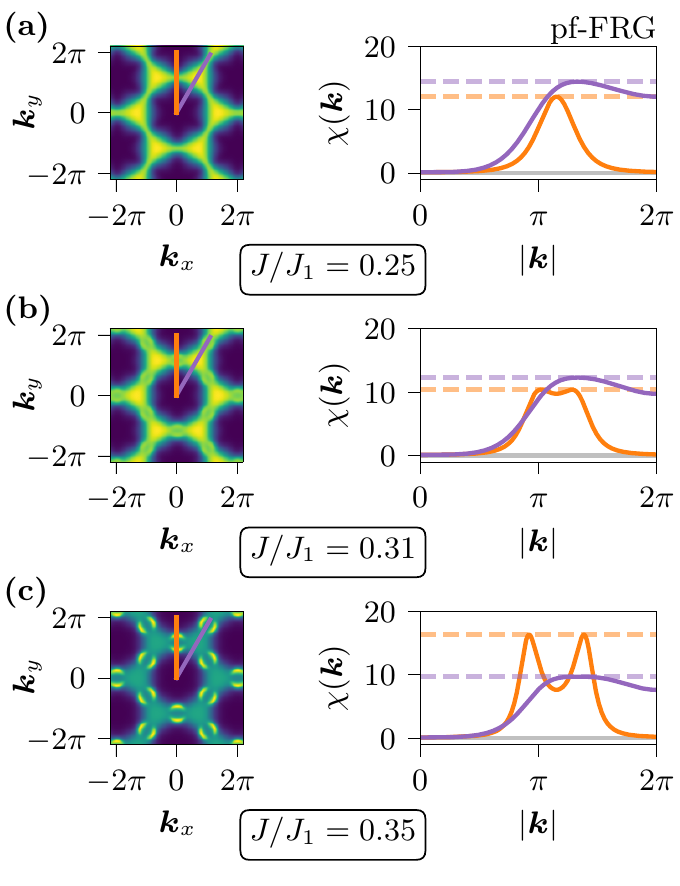}
    \caption{\textbf{Pinch-point to half-moon transition} from pf-FRG data at $\Lambda / Z = 0.01$. We plot the structure factor in the $\boldsymbol{k}_x - \boldsymbol{k}_y$ plane (left column) and along two distinct cuts through momentum space (right column). In the spin liquid phase (a), the maximum intensity is centered around the corners of the kagome Brillouin zone (magenta line), with subdominant peaks at the pinch-points (orange line). Approaching the half-moon phase (b) and (c), the pinch-points first flatten and split into two peaks and finally also carry the maximum intensity in the structure factor.}
    \label{fig:FRG_spectral}
\end{figure}


\subsection{Pinch-point to half-moon transition in pf-FRG}

To support the data regarding the pinch-point to half-moon transition presented in the main text, we explicitly present the pf-FRG data, from which the phase boundary was distilled. In Fig.~\ref{fig:FRG_spectral}, we plot structure factors close to the transition at $J /  J_1 \approx 0.33$ for $\Lambda / Z = 0.01$, both in the two dimensional $\boldsymbol{k}_x - \boldsymbol{k}_y$ plane as well as along two distinct momentum space cuts. In the spin-liquid phase (see panel (a) in Fig.~\ref{fig:FRG_spectral}), the structure factor peaks at the corners of the kagome Brillouin zone, as well as at the pinch-points, with more spectral weight distributed around the corners. Thus, the spectral measure $\Delta \chi_{\boldsymbol{k}}$, i.e the difference between the magenta and orange dashed line in the right column of Fig.~\ref{fig:FRG_spectral}, is positive. Around $J / J_1 \approx 0.31$, the peaks at the pinch-points flatten and give rise to two peaks (the half-moons), yet $\Delta \chi_{\boldsymbol{k}} > 0$ holds. Only at larger $J / J_1$, shown, e.g. in panel (c) of Fig.~\ref{fig:FRG_spectral}, the spectral measure changes sign, and the half-moons indeed pose the most distinct feature in the structure factor.


\subsection{Half-moon to star transition in pf-FRG}
\label{appendix:stars_pffrg}

\begin{figure}
    \centering
    \includegraphics[width=\columnwidth]{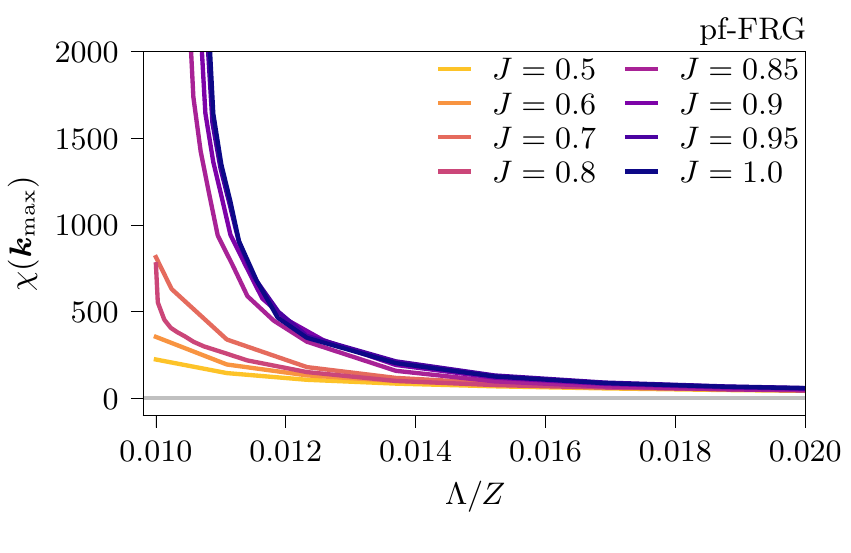}
    \caption{\textbf{pf-FRG flows of the structure factor} at the momenta with maximum intensity. For $J / J_1 < 0.8$ the flows are featureless and can be continued down to the smallest considered energy scale $\Lambda / Z = 0.01$. For $J / J_1 = 0.8$ the flow shows a sharp upturn at the lowest values of $\Lambda / Z$, which evolves into a divergence for $J / J_1 > 0.8$, signalling the onset of long-range magnetic order.}
    \label{fig:FRG_flows}
\end{figure}

\begin{figure}
    \centering
    \includegraphics[width=\columnwidth]{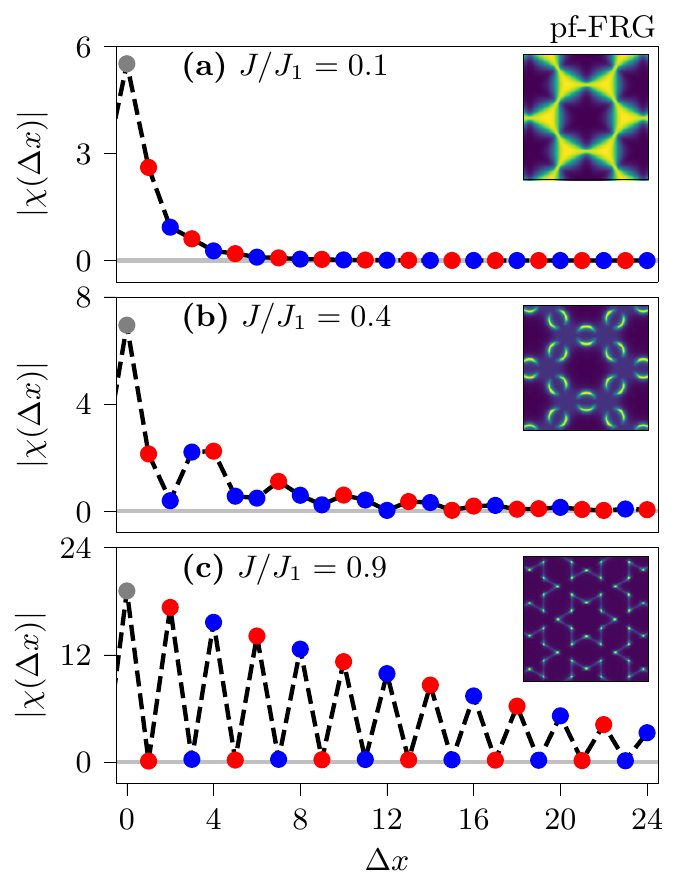}
    \caption{\textbf{Spin-spin correlations along the $\boldsymbol{a}_1$ direction of the kagome lattice} extracted from pf-FRG vertices at $\Lambda / Z \approx 0.011$ deep in the three different phases. Here blue (red) markers denote (anti-) ferromagnetic correlations to the reference site marked in grey. In (a) the spin liquid and (b) the half-moon phase, correlations decay particularly fast, and beyond $6 \sim 10$ bonds away from the origin their magnitude become negligible. In the star phase (c), however, magnetic correlations spread over the whole range of the lattice considered in the numerical calculations, explaining the observed flow breakdowns.}
    \label{fig:FRG_corrs}
\end{figure}

In contrast to the transition from the spin liquid (pinch-point) to the VBC (half-moon) phase, which could easily be identified in pf-FRG calculations by measuring the half-moon radius and the spectral parameter $\Delta \chi_{\boldsymbol{k}}$, determining the transition from the non-magnetic VBC to the magnetic collinear phase turns out to be more difficult. All other numerical approaches employed here consistently predict a finite magnetization around $J / J_1 \approx 0.5$, yet, the pf-FRG flows show no sign of a flow breakdown at this point (see Fig.~\ref{fig:FRG_flows}). Here, magnetic order sets in at larger couplings $J / J_1 \geq 0.8$ and only for extremely small cutoffs $\Lambda / Z \gtrsim 0.011$, close to the lower limit $\Lambda / Z = 0.01$ which is still numerically feasible. Probing the real-space correlations $\chi_{0 \ \Delta x \boldsymbol{a}_1} = \chi(\Delta x)$ along the $\boldsymbol{a}_1 = (1, 0)$ direction (i.e. along one axis of the kagome lattice), we indeed find fairly long-range correlations extending over the whole $L = 24$ real space cluster considered in the numerical simulations. In the spin liquid and half-moon phase, in contrast, correlations decay more rapidly and already for few bonds away from the origin, their magnitude is strongly diminished (see Fig.~\ref{fig:FRG_corrs}). The discrepancy in the precise location of the phase boundary could be related to the fulfillment of the half-filling constraint in pf-FRG. After all, it is only enforced on average and there may still exist fluctuations which populate unphysical, i.e. non-magnetic pseudofermion states~\cite{thoenniss2020}. These might be responsible for impeding the formation of a clear divergence of the RG flow already at smaller values of $J / J_1$. Furthermore, we cannot rule out a scenario in which the critical scale lies below the numerical threshold $\Lambda / Z = 0.01$.


\subsection{Dimer response from pf-FRG}

While the order parameter corresponding to a VBC state is of order $S^4$ and would require higher vertex functions that are out of reach for the pf-FRG, a qualitative picture of a system's tendency to select a particular dimer pattern may still be obtained.
To achieve this, the unit cell needs to be enlarged so that translational symmetry is broken by slightly increasing the strength of dimerized bonds while weakening the others, i.e. $J_1 \rightarrow J_1 \pm \delta$ with $\delta = 0.01 J_1$~\cite{Hering-2022,Iqbal-2016_nem,Keles2021}. Defining the equal time, real-space spin-spin correlation along such a strengthened dimer bond as $\langle S^z_i S^z_j \rangle_+$ and a completely unperturbed (i.e. $\delta = 0$) reference value $\langle S^z_i S^z_j \rangle_0$, we may define the dimer response as
\begin{equation}
    \chi_d = \frac{J_1}{\delta}\times \frac{\langle S^z_i S^z_j \rangle_+ - \langle S^z_i S^z_j \rangle_0 }{ \langle S^z_i S^z_j \rangle_0} \text{.} \label{eq:DimerResponse}
\end{equation}
Note that this definition requires the evaluation of two separate FRG runs to compute $\langle S^z_i S^z_j \rangle_0$ and $\langle S^z_i S^z_j \rangle_+$.
From pf-FRG, equal-time correlators can be computed as $\langle S^z_i S^z_j \rangle \equiv \langle S^z_i S^z_j \rangle(t=0) = \int d\nu \chi_{ij}(\nu)$.

\begin{figure}
    \centering
    \includegraphics[width=\columnwidth]{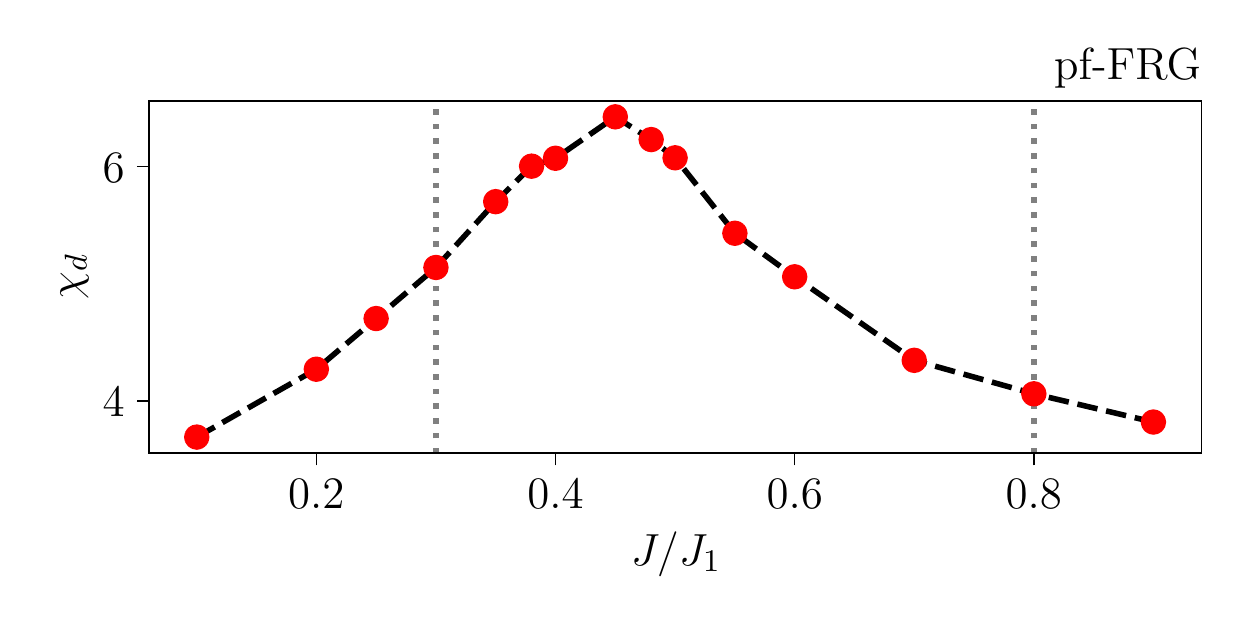}
    \caption{{\bf pf-FRG dimer response for the pinwheel VBC pattern}  in Fig.~\ref{fig:fig_1} of the main text as a function of $J/J_{1}$.
	The dimer response is calculated via Eq.~\eqref{eq:DimerResponse}. The phase boundaries shown as grey dotted lines are obtained as described in the sections above.}
    \label{fig:DimerResponse}
\end{figure}

Fig.~\ref{fig:DimerResponse} shows the response obtained for the pinwheel VBC pattern displayed in Fig.~\ref{fig:fig_1}, where thick red bonds are strengthened and thin bonds are weakened by $\delta$: In the QSL phase, we observe a relatively small value of the dimer response which rises steadily towards the VBC phase up until a distinct maximum at $J_2 \approx 0.45$ after which it decreases once more. This is in good agreement with the phase diagram presented in Fig.~\ref{fig:fig_1} of the main text.


\section{Many-variable wave function (mVMC)}
\label{appendix:mVMC}

The many-variable variational Monte Carlo (mVMC) method can be successfully used in studies of strongly correlated spin and electronic systems~\cite{PhysRevB.90.115137,casula2004correlated}. In particular, the method can be applied to distinguish between quantum spin liquid and valence bond solid phases, such as in the case of the $J_1$-$J_2$ Heisenberg model on the square lattice~\cite{doi:10.7566/JPSJ.84.024720,nomura2020dirac}. In this work, we employ the mVMC implementation from Ref.~\cite{misawa2019mvmc,doi:10.1143/JPSJ.77.114701}. The construction of the variational states relies on the Abrikosov fermion representation of spin degrees of freedom, as given in Eq.~\eqref{eq:partons}.

Inspired by the Anderson resonating valence-bond wave function, the mVMC ansatz has the form
\begin{equation}\label{eq:phipair}
    |\phi_{\mbox{\footnotesize pair}} \rangle = \hat{\mathcal{P}}^{\infty}_{\mbox{\footnotesize G}} \exp \left(\sum\limits_{i, j} F_{i,j} \hat f^{\dagger}_{i, \uparrow} \hat f^{\dagger}_{j, \downarrow} \right) |0 \rangle,
\end{equation}
where single occupation is ensured by the Gutzwiller projector
\begin{equation}\label{eq:gutzwiller}
    \hat{\mathcal{P}}^{\infty}_{\mbox{\footnotesize G}}=\prod_i (f^\dagger_{i,\uparrow}f_{i,\uparrow}-f^\dagger_{i,\downarrow}f_{i,\downarrow})^2,
\end{equation}
which maps the fermionic Hilbert space to the original Hilbert space of spin operators. The wave-function value $\langle \boldsymbol{\sigma} |\phi_{\mbox{\footnotesize pair}} \rangle $ of a specific spin configuration $|\boldsymbol{\sigma}\rangle$ is evaluated using the Slater determinant of the matrix with elements $F_{i,\,j}$. Here, $\boldsymbol{\sigma}$ represents a string of $\pm 1$, which, for each lattice site,  stands for the respective spin eigenstate in the $S^z$ basis. The parameters $F_{i,\,j}$ are optimised using the stochastic reconfiguration technique~\cite{sorella_green_1998}, which can be seen as a way of performing imaginary-time evolution in the variational parameters manifold~\cite{becca_quantum_2017,carleo2017solving}.

\begin{figure}
    \centering
    \includegraphics[width=\columnwidth]{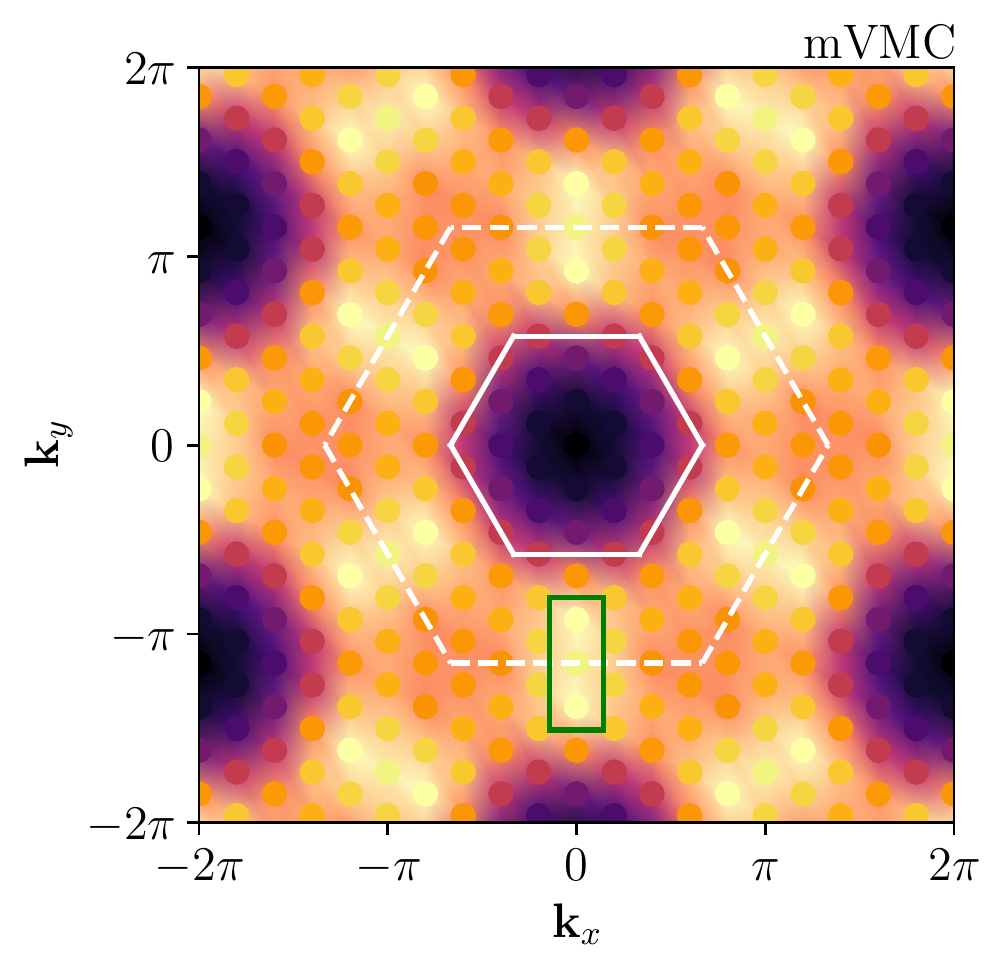}
    \caption{\textbf{Static (equal-time) spin structure factor} at $J/J_1=0.4$ as computed by mVMC. The color plot shows the isotropic structure factor $\chi(\mathbf{k})$ [Eq.~\eqref{eq:struct_fact}] in the $\boldsymbol{k}_x - \boldsymbol{k}_y$ plane. The results have been obtained on a $3 \times 8\times 8$ finite cluster. The white hexagon with solid (dashed) lines delimits the first (extended) Brillouin zone. The green box highlights the ``half-moon'' feature of the correlation pattern.}
    \label{fig:magnetic_2d}
\end{figure}

To improve the accuracy of the variational wave functions, we employ quantum-number projections. The point-group symmetry $\hat G$ is enforced by applying its generators until the symmetry orbit is exhausted
\begin{equation}
    |\Psi_\xi \rangle = \hat P |\Psi \rangle =  \sum\limits_n \xi^n \hat G^n |\Psi \rangle,
    \label{eq:q_projection}
\end{equation}
where $\xi$ is the desired projection quantum number and $|\Psi_\xi \rangle$ the resulting symmetrized state. The projection onto the total spin $S$ is performed by superposing the $SU(2)$--rotated wave functions~\cite{doi:10.1143/JPSJ.77.114701}. In this work, for systems with more than 36 sites, we partially impose translational symmetry directly on the variational parameters $F_{i,\,j}$. Namely, we introduce translational symmetry modulo $2 \times 2$ unit cells sublattice structure and enforce the $2 \times 2$ translations and the point-group symmetries using Eq.\,\eqref{eq:q_projection}. The resulting procedure amounts into $2 \times 2 \times 3^2 \times L^2$ variational parameters with $L$ being the number of unit cells in each lattice direction. Such partial translational symmetry imposition is a reasonable compromise between the ability to express complicated wave function and the required time to optimize the wave function.

Magnetic properties of variational wave functions can be assessed by computing the structure factor $\chi(\mathbf{k})$ as the equal-time momentum-resolved spin-spin correlation function
\begin{equation}\label{eq:struct_fact}
    \chi(\mathbf{k})=\frac{1}{3L^2}\sum_{i,\,j} e^{i \mathbf{k}\cdot (\mathbf{r}_i-\mathbf{r}_j)} \langle \hat{\mathbf{S}}_i \cdot \hat{\mathbf{S}}_j\rangle,
\end{equation}
where $\mathbf{r}_i$ indicates the position of the lattice site $i$ including sublattice displacement. In Fig.\,\ref{fig:magnetic_2d}, we present the spin structure factor at $J/J_1=0.4$ as computed by mVMC.

\begin{figure}[t]
    \centering
      \includegraphics[width=1.\textwidth]{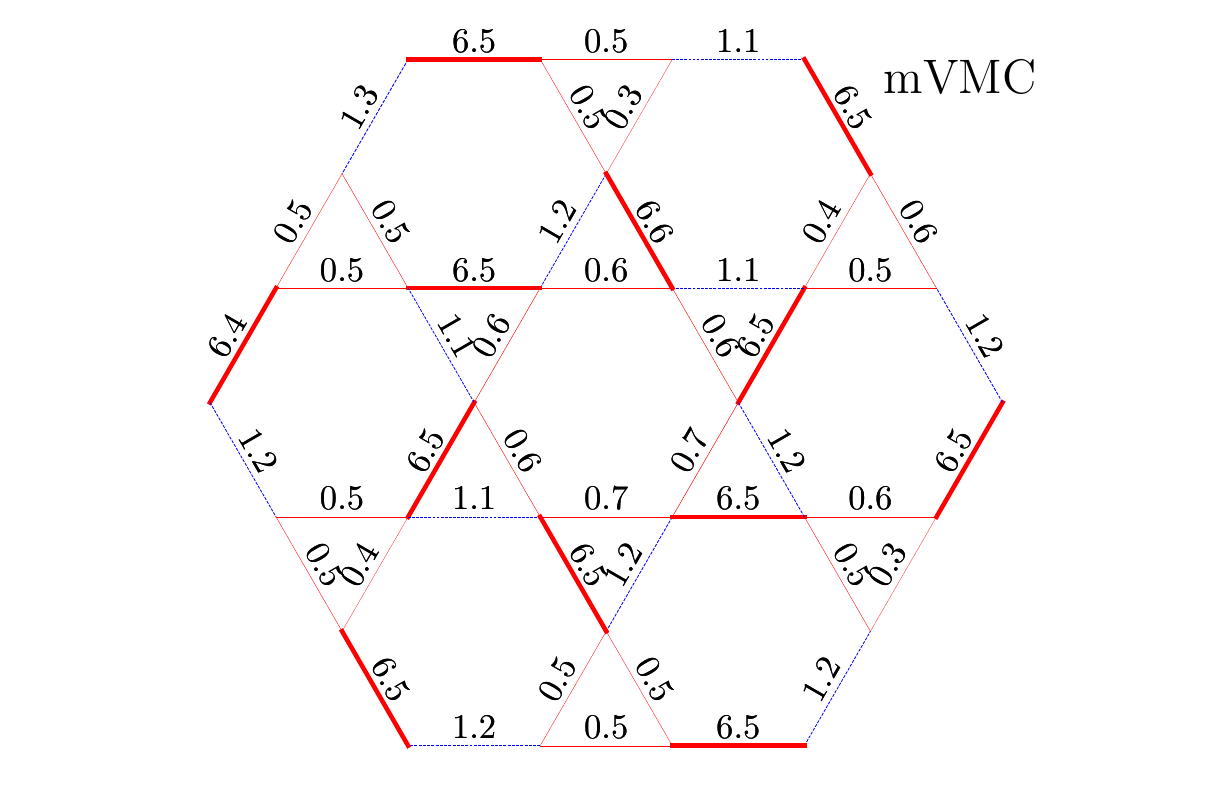}
    \caption{{\bf Real-space correlations pattern} $\langle \hat{D}_a \hat{D}_b \rangle - \langle \hat{D}_a \rangle \langle \hat{D}_b \rangle$ measured within mVMC on the $3\times 8 \times 8$ kagome lattice for $J / J_1 = 0.4$. Here $\hat{D}_a$ is the dimer operator placed on the ``base'' bond (in a distant unit cell) and $\hat{D}_b$ is the dimer operator on other bonds. Red (blue) color in the figure represents positive (negative) values of the correlator, while its absolute magnitude (multiplied by 100) is marked near each bond. The correlations were measured on a non-symmetrized mVMC wave function for which the pinwheel dimer pattern is more pronounced.}
    \label{fig:dimer_real_space_pattern}
\end{figure}

In a non-magnetic phase, the properties of the wave function are assessed by measuring the dimer-dimer correlation function $\chi^D_{b,\,b'} = \langle \hat{D}_b \hat{D}_{b'} \rangle - \langle \hat{D}_b \rangle \langle \hat{D}_{b'} \rangle$ for all pairs of bonds in the system, $0 \leqslant b,\,b' < N_{\mbox{\small bonds}}$, where $\hat{D}_b=\mathbf{\hat S}_{i} \cdot \mathbf{\hat S}_{j}$, with $i,\,j$ being sites at ends of the bond $b$. In Fig.~\ref{fig:dimer_real_space_pattern}, we show the dimer-dimer correlations between the base bond (located in a distant unit cell) and other bonds. To carry out a quantitative assessment of the VBC character of the ground state, we need to define suitable scalar order parameters to perform an infinite-volume extrapolation of the dimer order. Thus, we regard $\chi^D_{b,\,b'}$ as a matrix in the bond indices and we diagonalize it; the resulting set of eigenvalues/eigenvectors pairs (${\lambda, \ A^\lambda_b}$) is used to define the operators $\hat{\mathcal{O}}_\lambda = \sum_b A^\lambda_b \hat D_b$, each of them corresponding to a certain momentum and irreducible representation of the lattice point group. The tendency to establish a finite expectation value of one of these operators, and thus spontaneously break the corresponding lattice symmetry, is measured by the susceptibility ${\chi_{\hat{\mathcal{O}}_\lambda} = \langle \hat{\mathcal{O}}_\lambda^{\dagger} \hat{\mathcal{O}}_\lambda \rangle -
\langle \hat{\mathcal{O}}_\lambda^{\dagger} \rangle\langle \hat{\mathcal{O}}_\lambda \rangle=\lambda}$ extrapolated to the thermodynamic limit~\cite{astrakhantsev2021brokensymmetry}.

\begin{figure}[t]
    \centering
    \begin{tikzpicture}
    \node[inner sep=0pt] at (0,0)    {\includegraphics[width=1.\textwidth]{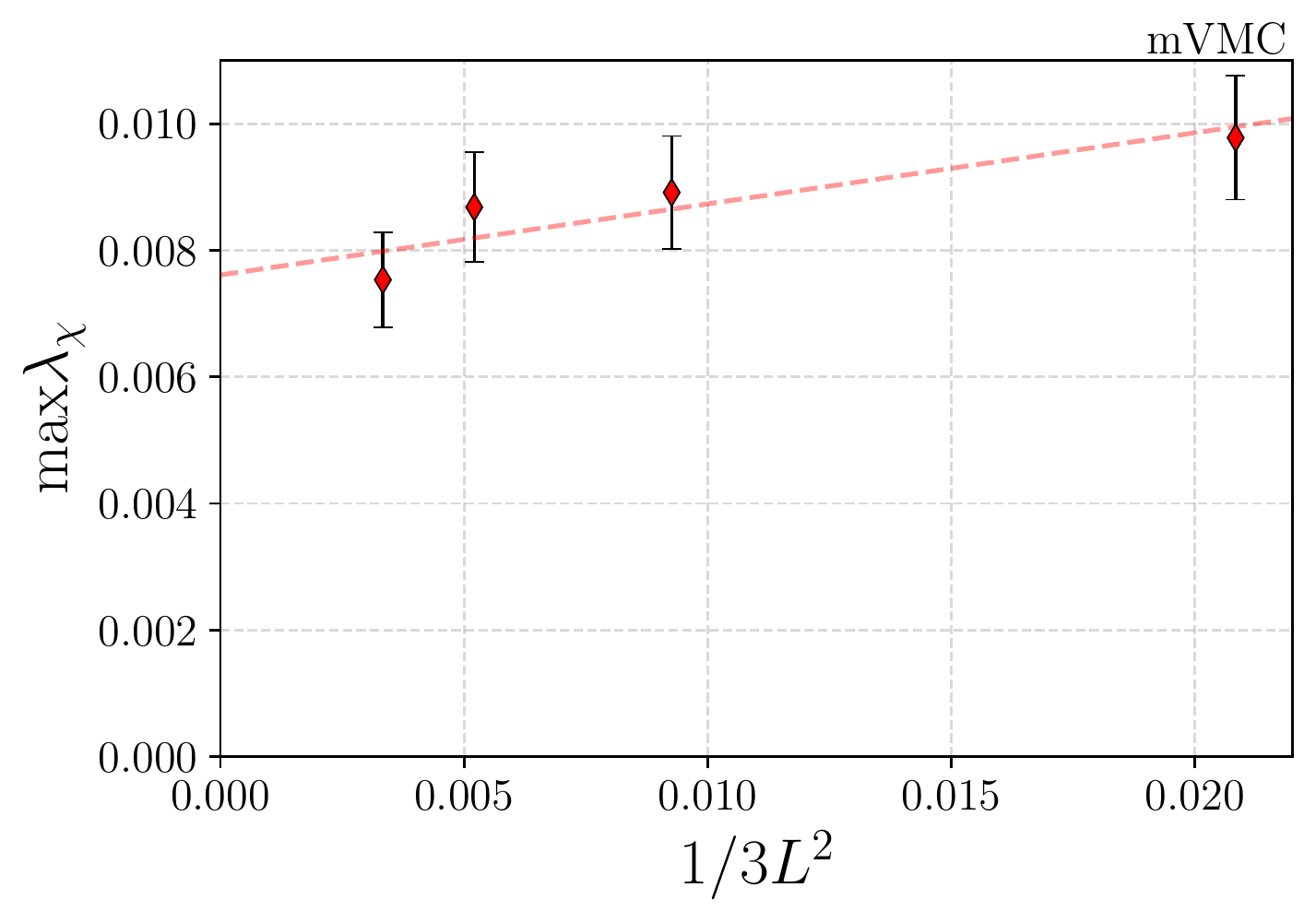}};
    \node[inner sep=0pt] at (2.0, 0)    {\includegraphics[width=0.6\textwidth]{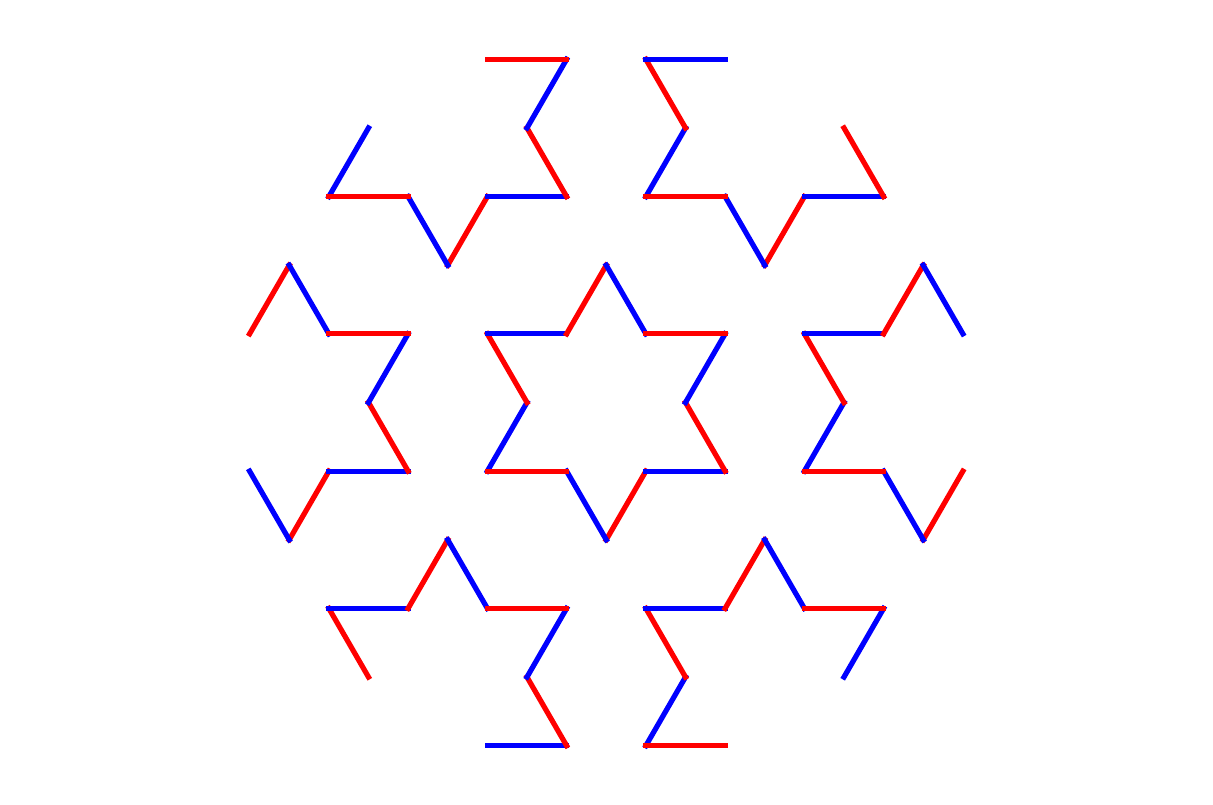}};
    \end{tikzpicture}
    \caption{{\bf Extrapolation of the maximum eigenvalue of the $\chi^D_{ij}$ dimer-dimer correlation matrix} at $J / J_1 = 0.4$ as the function of inverse lattice volume. The leading eigenvalue is triply-degenerate, while the other eigenvalues are an order of magnitude smaller. The inset shows an equal-weight superposition of the three basis functions of the dominant irreducible representation, all three connected by $2 \pi / 3$--rotations.}
    \label{fig:dimer_correlations_pattern}
\end{figure}

Following this procedure, we obtain leading eigenvalues and eigenstates of the $\chi^D_{b,\,b'}$ matrix at $J/J_1 = 0.2,\,0.3,\,0.4$ on finite-size lattices with $L = 4,\,6,\,8$ and $10$. In Fig.\,\ref{fig:dimer_correlations_pattern}, we show an equal-weight superposition of the three degenerate leading eigenstates at the $M$--points. Other eigenvalues are an order of magnitude smaller, and are thus not shown. We extrapolate the corresponding susceptibility to the thermodynamic limit and obtain, for $J/J_1 = 0.4$, non-vanishing susceptibility extrapolation of $7.6(3) \times 10^{-3}.$ This signals presence of symmetry breaking through establishment of a dimer order at $J/J_1 = 0.4$. Similar extrapolations at $J/J_1 = 0.2,\,0.25,\,0.3$ and $0.35$ yield $3.8(2) \times 10^{-3}$, $4.1(2) \times 10^{-3}$, $5.1(3) \times 10^{-3}$ and $6.9(2) \times 10^{-3}$, respectively. By fitting the susceptibility dependence on $J/J_1$ with a hyperbolic tangent ansatz, we estimate the inflection point to be at $J/J_1 = 0.32(3)$. This provides an estimate of the transition point from the QSL to the VBC phase. We emphasize that the dimer-dimer susceptibility within mVMC remains finite in the QSL phase. This is related to the fact that the mVMC wave function cannot efficiently express the ground state of the $U(1)$ DSL phase, unlike the dimerized VBC case. This is confirmed by the comparison of the mVMC variational energy to the one of the DMRG approach in the two phases. 


\begin{figure*}
    \centering
    \includegraphics[width=0.33\columnwidth]{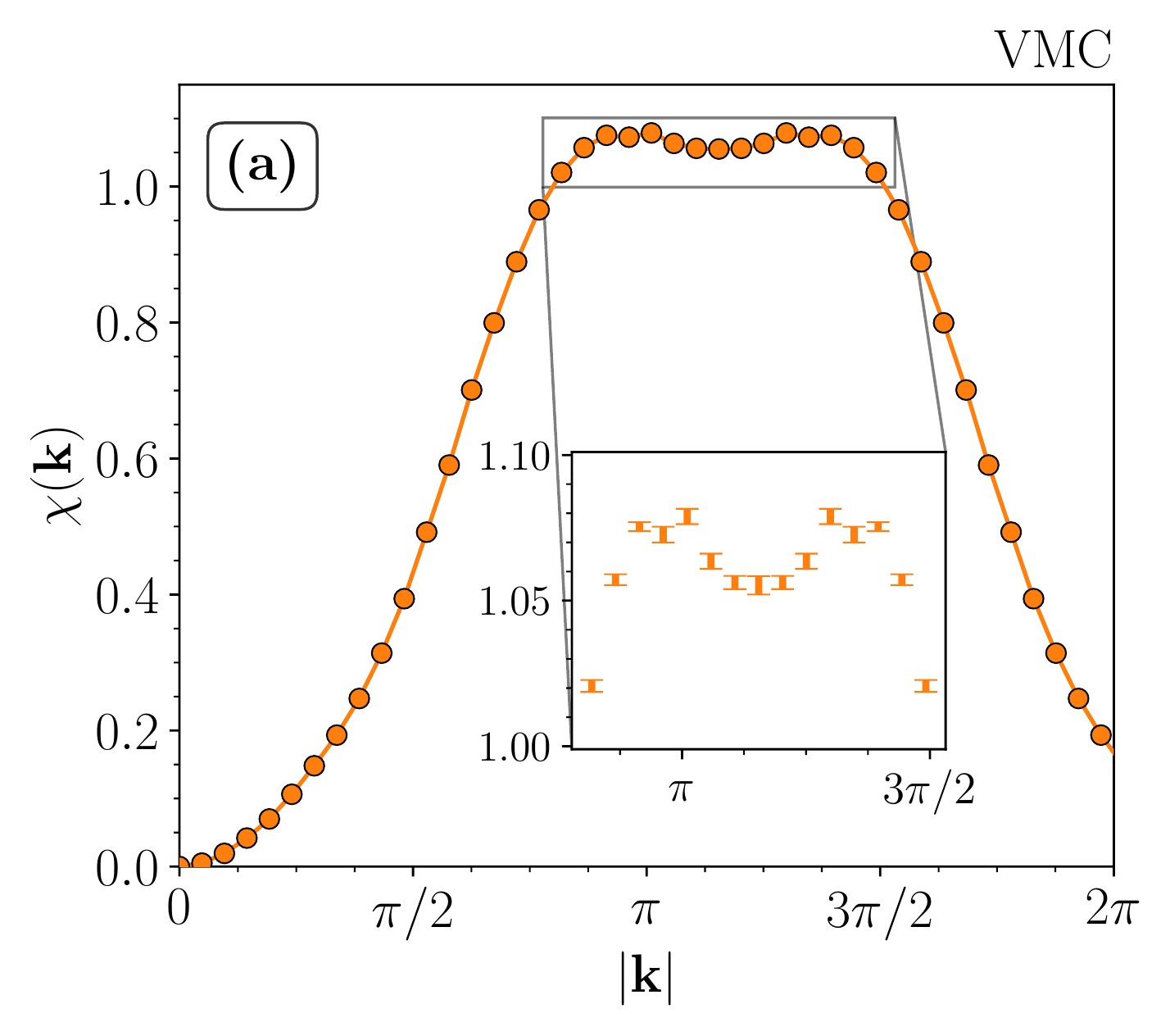}
    \includegraphics[width=0.33\columnwidth]{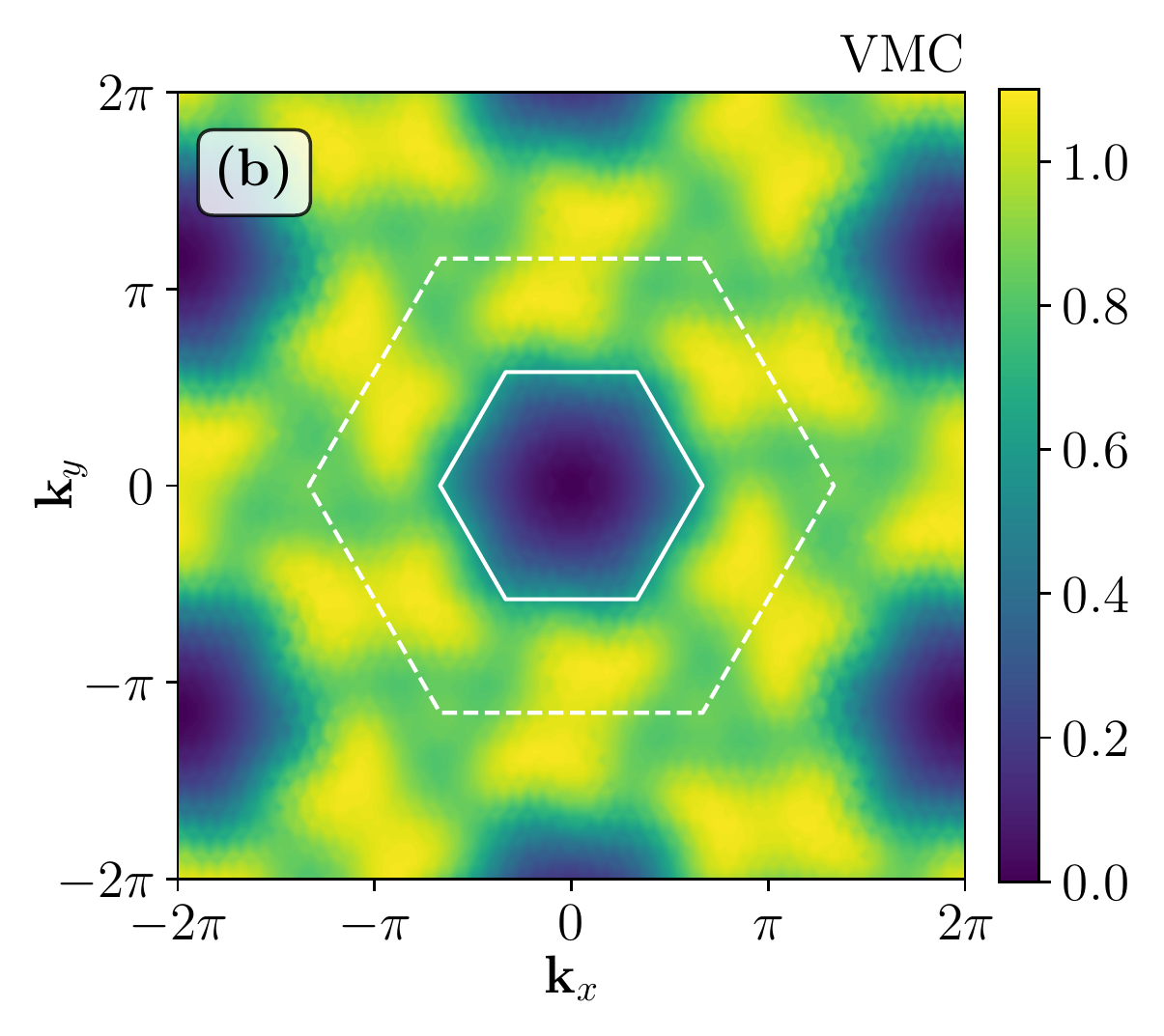}
    \includegraphics[width=0.33\columnwidth]{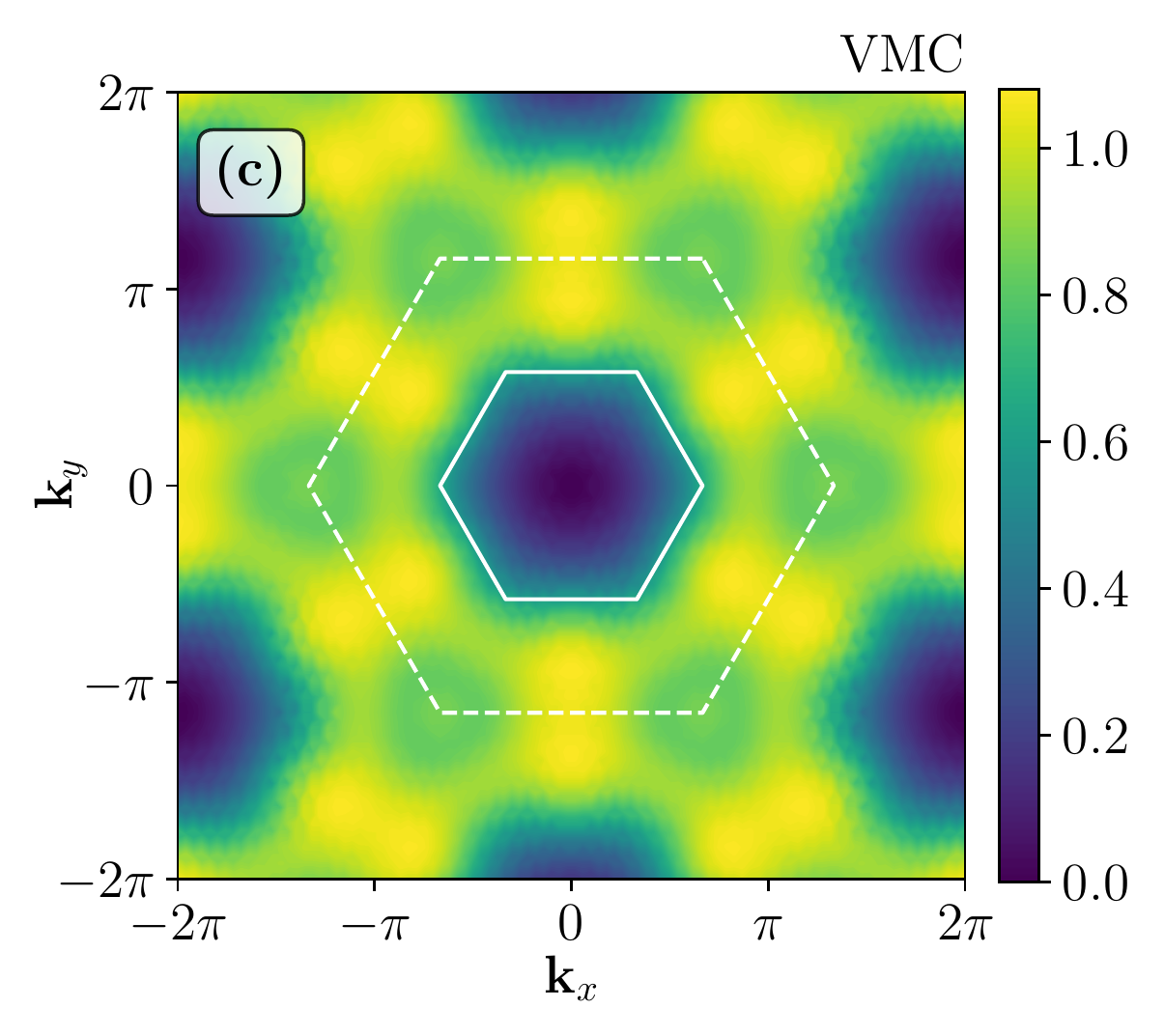}
    \caption{\label{fig:halfmoon_vmc} \textbf{Static (equal-time) spin structure factor of the VBC phase} as computed by VMC [Eq.~\eqref{eq:struct_fact}]. Results for $J/J_1=0.5$ are shown (on a $3 \times 24\times 24$ lattice). 
	In (a) we plot $\chi(\mathbf{k})$  as a function of $|\mathbf{k}|$, along the $\mathbf{k}_x=0$ cut 
		in momentum space shown in Fig.~\ref{fig:FRG_spectral} (orange vertical line). 
		The inset highlights the dip in the profile of $\chi(\mathbf{k})$  (with error bars) at the pinch-point positions, 
		which signals the appearance of half-moons. 
		The color plots in (b) and (c) show the value of $\chi(\mathbf{k})$  in the $\boldsymbol{k}_x - \boldsymbol{k}_y$ plane. 
		The results of (b) are obtained with the bare VBC wave function, which breaks the reflection symmetry of the kagome lattice. 
		In (c) we show symmetrized results for the structure factor, where the reflection symmetry is imposed \textit{a posteriori}. 
		The white hexagon with solid (dashed) lines delimits the first (extended) Brillouin zone.
		}
\end{figure*}

\section{Variational Monte Carlo (VMC)}

The variational Monte Carlo (VMC)~\cite{becca_quantum_2017} approach employed in this work shares several common aspects with the mVMC method introduced in the previous section. Both techniques rely on the Abrikosov fermion representation of spin operators, introduced in Eq.~\eqref{eq:partons}. Within this fermionic formulation, suitable variational states for the Heisenberg model are obtained by projecting a fermionic wave function to the spin Hilbert space. The projection, which enforces the single fermionic occupation of each lattice site, is achieved by means of a Gutzwiller-projector $\mathcal{P}^\infty_G$ [see Eq.\,\eqref{eq:gutzwiller}], and can be performed exactly by an appropriate Monte Carlo sampling~\cite{becca_quantum_2017}. The variational \textit{Ans\"atze} of VMC, discussed in this section, differ from those of the mVMC approach in the choice of the fermionic states to be projected. 

Within the VMC approach, the variational state is obtained by projecting a Slater determinant, $|\Phi_{\rm 0}\rangle$, which is the ground state of an auxiliary quadratic Hamiltonian
\begin{equation}\label{eq:h0_vmc}
	\mathcal{H}_{\rm 0}
	=\sum_{i,j} t_{ij}(f_{i,\uparrow}^{\dagger}f_{j,\uparrow}^{\phantom\dagger}
	+ f_{i,\downarrow}^{\dagger}f_{j,\downarrow}^{\phantom\dagger}) 
	+  \sum_{i}\sum_{\mu=x,y,z} h^\mu_i S^\mu_i \,.
\end{equation}
The parameters $t_{ij}$ (hoppings) and $h^\mu_i$ (fictitious magnetic field) of $\mathcal{H}_{\rm 0}$ are optimized in order to minimize the variational energy of the projected state \footnote{Additional superconducting pairing terms can be included in the auxiliary Hamiltonian of Eq.~\eqref{eq:h0_vmc}, but for the model under consideration they do not provide any energy gain in the thermodynamic limit.}. The complete expression for the variational wave function is
\begin{equation}
    |\Psi_{\rm var}\rangle= \mathcal{J} \mathcal{P}^\infty_G |\Phi_{\rm 0}\rangle \,,
\end{equation}
where, in addition to the projected Slater determinant, we have included the long-range spin-spin Jastrow factor~\cite{becca_quantum_2017}
\begin{equation}\label{eq:jastrow}
    \mathcal{J}=\exp\Big(\sum_{i,j} v_{i,j} S_i^z S_j^z\Big) \,.
\end{equation}
The pseudopotential parameters $v_{i,j}$ are assumed to be translationally invariant, and numerically optimized along with the fermionic parameters $t_{ij}$ and $h^\mu_i$. The optimization of the variational wave function is achieved through the stochastic reconfiguration method~\cite{sorella_green_1998,becca_quantum_2017,carleo2017solving}

\subsection{Spin liquid to pinwheel VBC transition in VMC}

For small values of the ratio $J/J_{1}$, the optimal variational wave function for the model is the U(1) DSL~\cite{Ran-2007,Iqbal-2013}.
Upon increasing $J/J_{1}$, the system undergoes a phase transition to the pinwheel VBC at $J/J_{1}=0.26(1)$ (see Fig.~\ref{fig:first_transition} of the main text). The variational Ansatz for the VBC is obtained by considering a $2 \times 2$ enlarged unit cell of $12$ sites, which can accommodate the pinwheel pattern depicted in Fig.~\ref{fig:fig_1}~(b) of the main text. 
The variational parameters of the VBC Ansatz are the inequivalent hoppings within the enlarged unit cell, from first- to third-neighbors (the latter ones being limited to the $J_{3a}$-bonds). The number of independent hopping parameters is reduced from $72$ to $12$ by applying the $C_6$ rotational symmetry of the kagome lattice. Finally, an underlying sign structure for the hoppings is imposed, to reproduce the flux pattern of the U(1) DSL (similarly to the approach used in Ref.~\cite{Iqbal_2012}). For this reason, the VBC wave function can be regarded as an instability of the DSL state. We find that the energy of the VBC state is lower than the one of the DSL for $J/J_{1}>0.26$, signalling the transition to the pinwheel VBC phase. 
Deep inside the VBC phase, the static structure factor $\chi(\mathbf{k})$ displays signatures of half-moons, as shown in Fig.~\ref{fig:halfmoon_vmc}.

\subsection{Pinwheel VBC to collinear magnetic order transition in VMC}

\begin{figure}
    \centering
    \includegraphics[width=0.75\columnwidth]{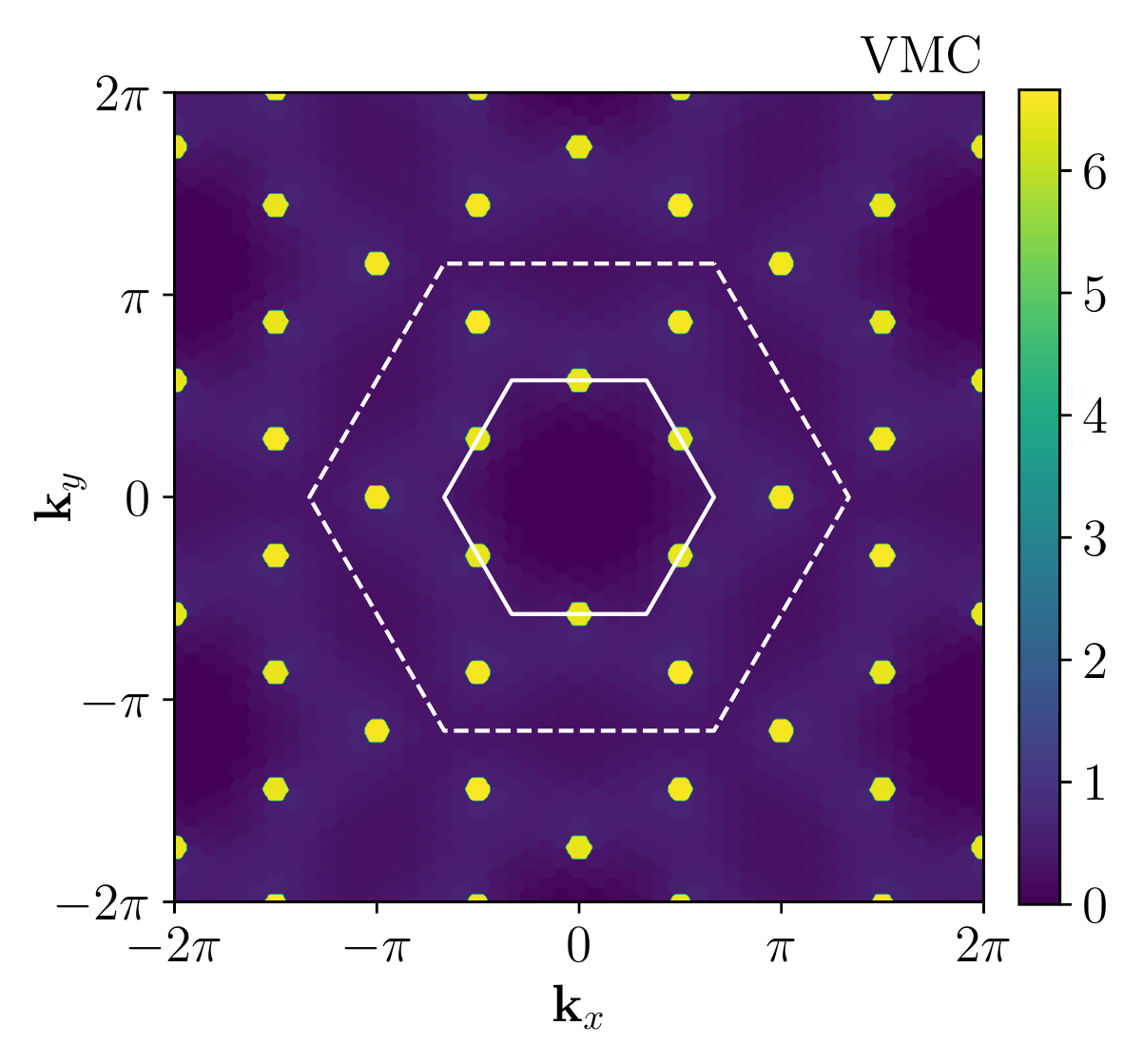}
    \caption{\textbf{Static (equal-time) spin structure factor of the collinear ordered phase} at $J/J_1=0.6$ as computed by VMC. The color plot shows the isotropic structure factor $\chi(\mathbf{k})$  in the $\boldsymbol{k}_x - \boldsymbol{k}_y$ plane. The results have been obtained on a $3 \times 12\times 12$ finite cluster. The white hexagon with solid (dashed) lines delimits the first (extended) Brillouin zone.} 
    \label{fig:magnetic_2d_octahedral}
\end{figure}

\begin{figure}
    \centering
    \includegraphics[width=\columnwidth]{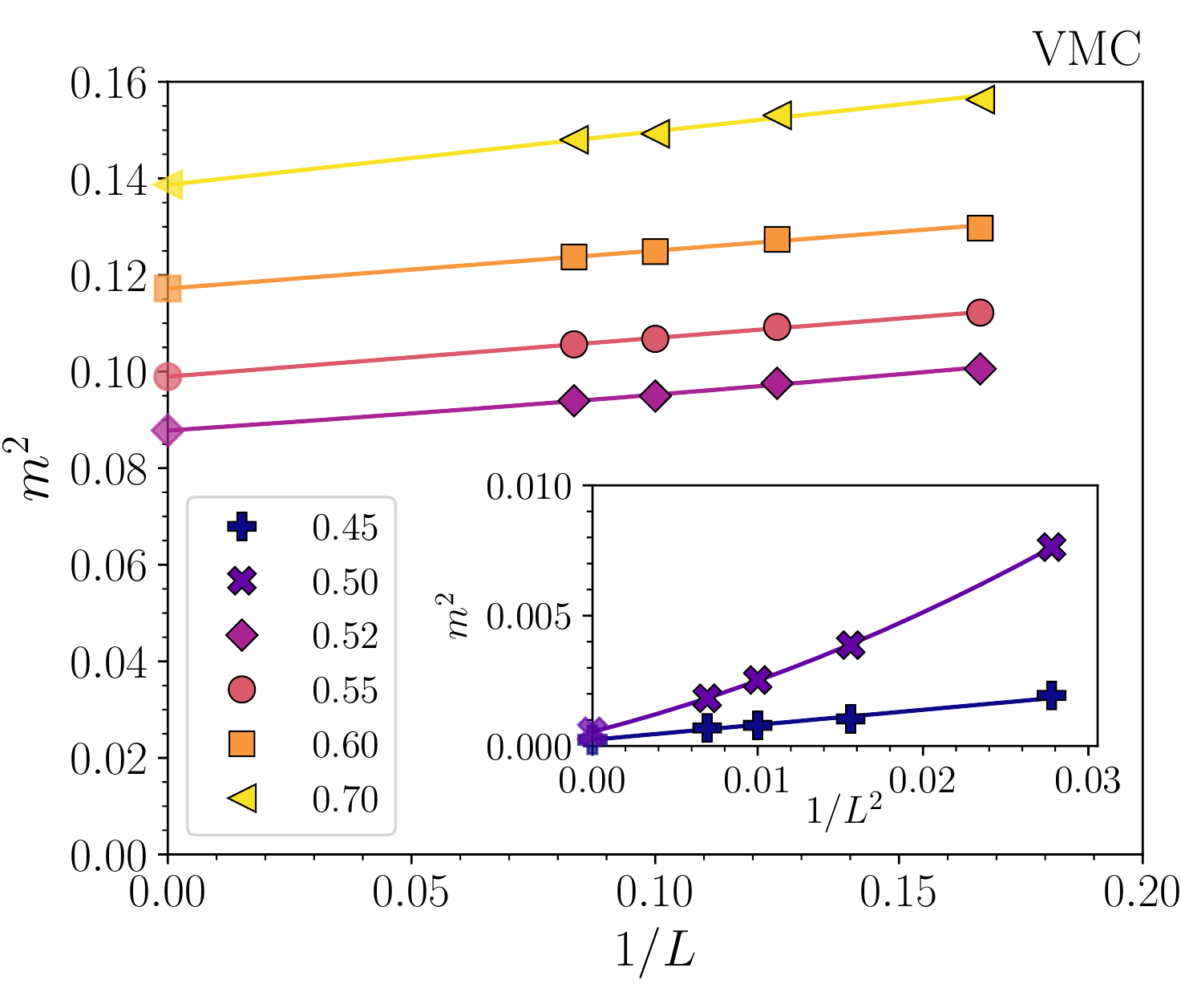}
    \caption{\textbf{Finite-size scalings of the square of the sublattice magnetization} $m^2$ for the collinear spin order showing its behavior at the VBC-magnetic order transition. We employed $3\times L\times L$ clusters with $L=6,8,10,12$. The values of $J/J_1$ are reported in the box in the lower-left corner. The inset shows the finite-size scaling of $m^2$ within the VBC regime, as a function of $1/L^{2}$. For $J/J_{1}>0.51$ (i.e., within the collinear magnetic phase), $m^2$ extrapolates to a non-zero value for $L\rightarrow \infty$. }
    \label{fig:m2_vmc_scaling}
\end{figure}

At $J/J_1=0.51(1)$, a phase transition from the pinwheel VBC to the magnetic phase with collinear order is observed. The auxiliary Hamiltonian $\mathcal{H}_0$ for the magnetic state features the same hopping parametrization of the VBC Ansatz, with the addition of a fictitious magnetic field $h_i^\mu$ which reproduces the collinear order sketched in the inset of Fig.~\ref{fig:mvmc_m2}. The fictitious collinear field $h_i^\mu$ is taken along the $S_x$ direction. Thus, the presence of the Jastrow factor, which is a function of $S_z$ operators, introduces transverse spin fluctuations on top of the ordered fermionic state. Although the variational parametrization allows for a continuous transition between the VBC and the collinear ordered states, the transition turns out to be of the first order. Indeed,
we detect the presence of two energy minima when optimizing the variational energy, i.e. an absolute minimum and a metastable state with higher energy. One of the minima corresponds to the VBC state, i.e., it is characterized by a vanishingly small magnetic field $h_i^\mu$ in the thermodynamic limit and a dimer pattern like the one of Fig.~\ref{fig:fig_1}~(b); the other minumum, instead, corresponds to the magnetically ordered phase. The relative positions of the two minima swap at $J/J_1=0.51(1)$, and magnetic order sets in for larger values of $J/J_1$. In the collinear ordered phase, the static structure factor shows the presence of Bragg peaks at the ordering vectors, as shown in Fig.~\ref{fig:magnetic_2d_octahedral} for $J/J_1=0.6$. The first-order nature of the VBC-collinear order transition is confirmed by the sudden jump of the sublattice magnetization, shown in Fig.~\ref{fig:magnetic_2d} of the main text for a $3\times 8 \times 8$ lattice. A finite-size scaling analysis of $m^2$ confirms the presence of an abrupt change at the phase boundary also in the thermodynamic limit (see Fig.~\ref{fig:m2_vmc_scaling}).\\


\section{Density Matrix Renormalization Group}

\begin{figure}[h]
    \centering
    \includegraphics[width=\columnwidth]{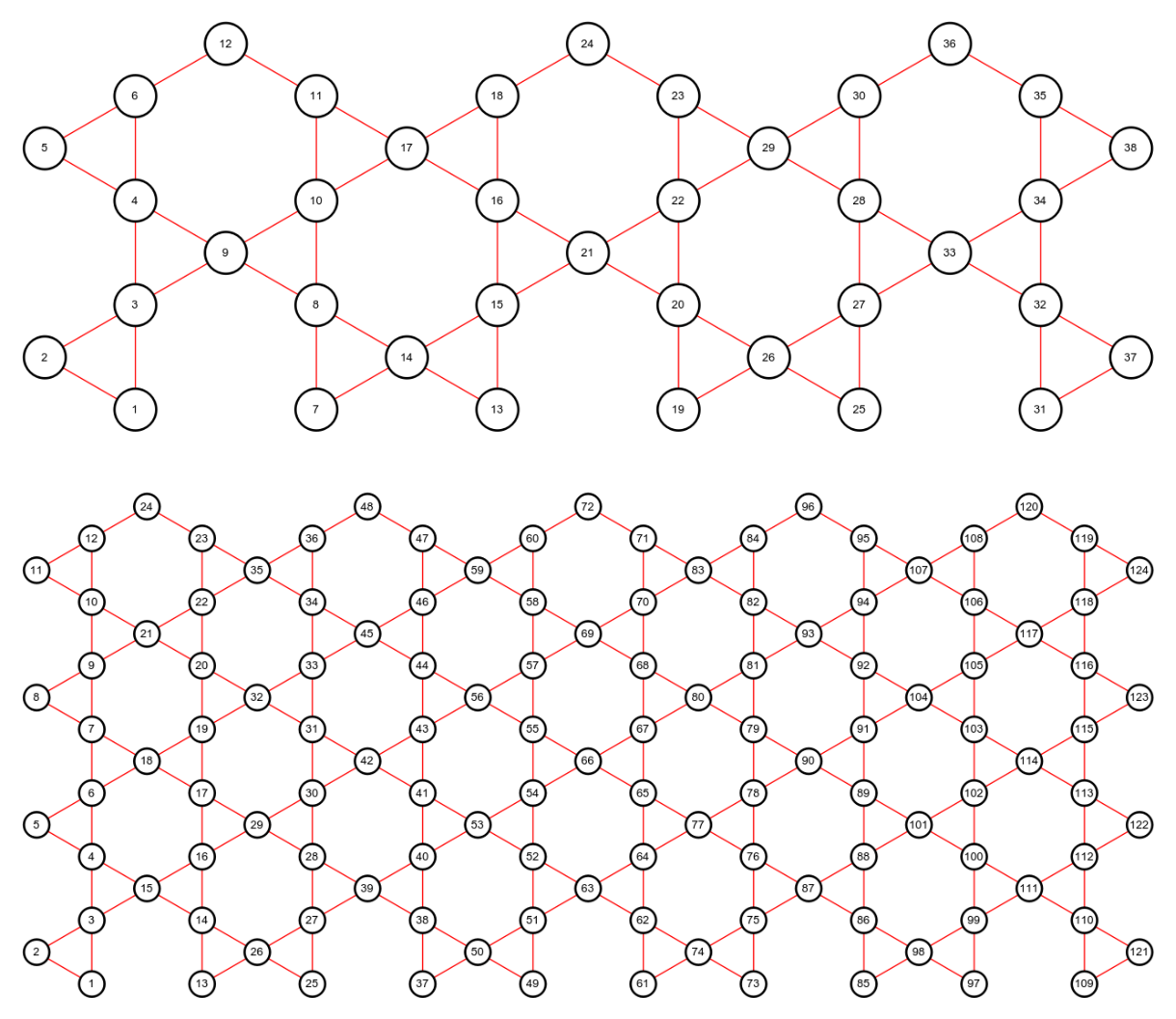}
    \caption{{\bf DMRG clusters.} 
    		The YC4-4 (38 sites) and YC8-8 (124 sites) spin tubes used for the DMRG calculations.}
    \label{fig:DMRG_clusters}
\end{figure}

Our density matrix renormalization group (DMRG) calculations are performed with the matrix product state (MPS) algorithm using the ITensor library~\cite{Fishman2020} on YC4-4 (38 sites) and YC8-8 (124 sites) spin tubes as illustrated in Fig.~\ref{fig:DMRG_clusters}, with 4 and 8 sites lying on the $y$-axis with a periodic boundary condition implemented along the $y$-axis. Along $x$-axis the system is open.The maximum bond dimension used for these calculations is $2048$. In general, for each DMRG run we are performing 12 full sweeps.  

\subsection{Spin liquid to pinwheel VBC transition in DMRG}

\begin{figure}[h!]
    \centering
    \includegraphics[width=\columnwidth]{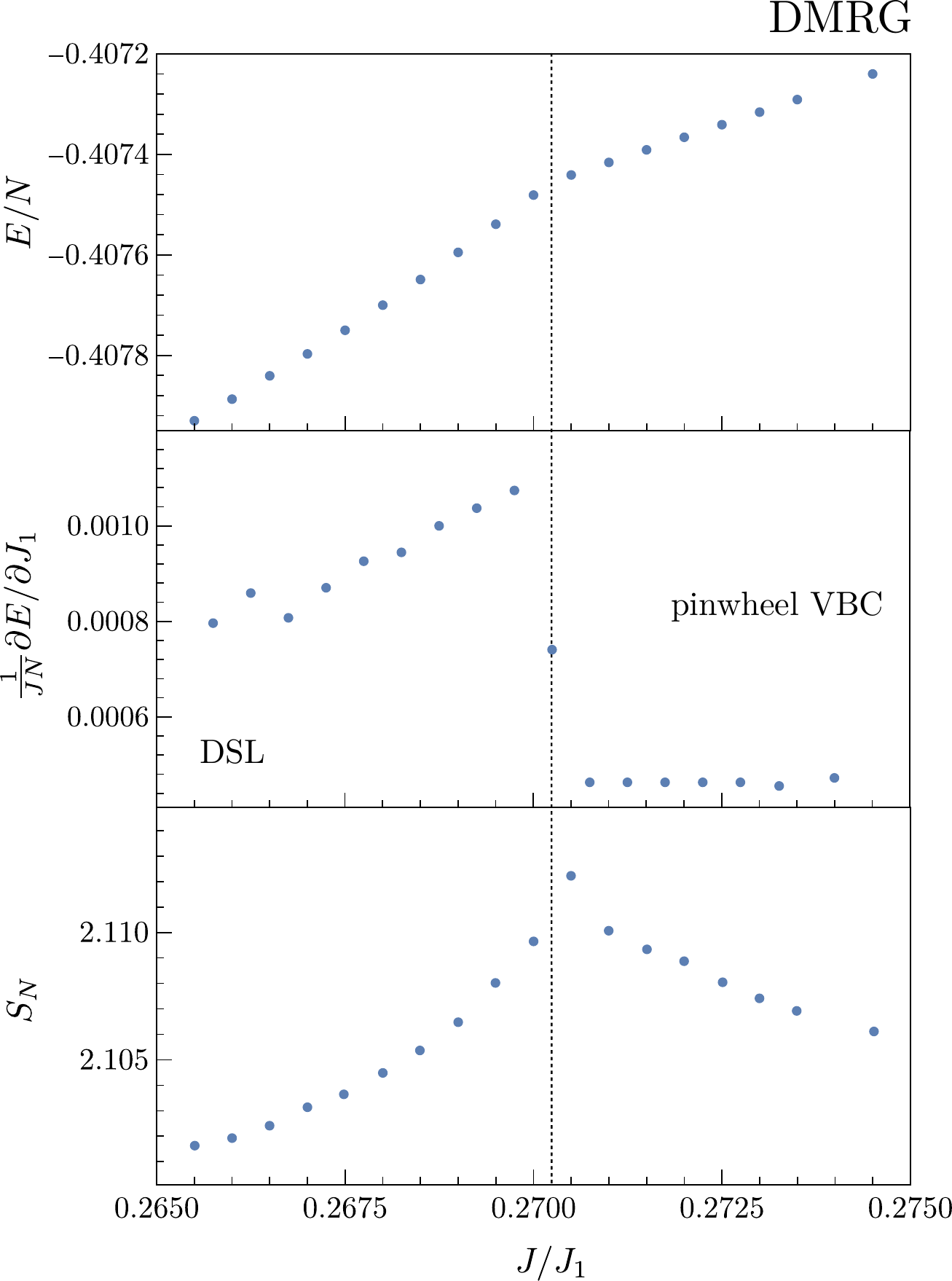}
    \caption{{\bf DMRG results for the spin liquid to pinwheel VBC transition.}
    		Top: The ground state energy as a function of $J/J_1$ and its derivative (middle panel).
		Bottom: The von Neumann entanglement entropy $S_N$ calculated across the central bond 
		using the matrix product ground state obtained via DMRG. 
		The kink in the ground state energy, leading to a step-function behavior in its derivative, and the sharply kinked,
		non-monotonous behavior of $S_N$ are all indicative of a first-order phase transition. 
		Its location is estimated at $J/J_1 \approx 0.27(1)$ indicated by the dashed line, 
		consistent with results from VMC calculations, see Fig.~\ref{fig:first_transition}(a).}
    \label{fig:entropy}
\end{figure}

In our DMRG calculations, we find $g_1$ is occurring at $J/J_1=0.27(1)$ which is signalled in (i) a discontinuity in the derivative of ground state energy with respect to $J$ as plotted in Fig.~\ref{fig:first_transition}(a) of the main text, and (ii) a sharp kink in the von Neumann entanglement entropy right at this transition as shown in Fig.~\ref{fig:entropy}.
Both signatures are consistent with a first-order transition.   

\clearpage


\end{document}